\documentclass[a4paper,11pt]{article}

\usepackage[DIV10]{typearea}
\usepackage[T1]{fontenc}
\usepackage[bookmarksopen,colorlinks]{hyperref}
\usepackage{graphicx}
\usepackage[latin1]{inputenc}
\usepackage{amssymb}
\usepackage{color}
\usepackage{textcomp}
\usepackage{slashed}
\usepackage{amsmath,bm}
\usepackage{cite}
\usepackage{axodraw}

\hypersetup{
   pdfstartview={FitH},
   pdfpagemode={UseOutlines}
}

\def\lsim{\:\raisebox{-0.5ex}{$\stackrel{\textstyle<}{\sim}$}\:}

\newcommand{\GEV}{\ensuremath{\,\textnormal{GeV}}}
\newcommand{\TEV}{\ensuremath{\,\textnormal{TeV}}}

\definecolor{bgreen}{RGB}{5,160,15}

\begin{document}

\date{\today}

\title{\Large 
Gravitino Dark Matter in Split Supersymmetry with Bilinear R-Parity Violation
}
\author{Giovanna Cottin$^1$, Marco A. D\'\i az$^2$, Mar\'\i a Jos\'e Guzm\'an$^3$ and Boris Panes$^4$  \\[2mm] 
{\small
${}^1${\it Cavendish Laboratory, University of Cambridge, J.J. Thomson Ave, Cambridge CB3 0HE, UK}}
\\
{\small
${}^2${\it Instituto de F\'\i sica, Universidad Cat\'olica de Chile, Av. Vicu\~na Mackenna 4860, Santiago, Chile}}
\\
{\small
${}^3${\it Instituto de Astronom\'\i a y F\'\i sica del Espacio, C.C. 67, Suc. 28, 1428, Buenos Aires, Argentina}}
\\
{\small
${}^4${\it Instituto de F\'\i sica, Universidade de S\~ao Paulo, R. do Mat\~ao 187, S\~ao Paulo, SP, 05508-900, Brazil}}
\\
{\small\tt gfc24@hep.phy.cam.ac.uk, mad@susy.fis.puc.cl,}
\\
{\small\tt mjguzman@iafe.uba.ar, bapanes@if.usp.br}
}

\maketitle

\begin{abstract}
In Split-SUSY with BRpV we show that the Gravitino DM solution is consistent with experimental evidence on its relic density and life time. We arrive at this conclusion by performing a complete numerical and algebraic study of the parameter space, including constraints from the recently determined Higgs mass, updated neutrino physics, and BBN constraints on NLSP decays. The Higgs mass requires a relatively low Split-SUSY mass scale, which is naturally smaller than usual values for reheating temperature, allowing the use of the standard expression for the relic density. We include restrictions from neutrino physics with three generations, and notice that the gravitino decay width depends on the atmospheric neutrino mass scale. We calculate the neutralino decay rate and find it consistent with BBN. We mention some implications on indirect DM searches.
\end{abstract}

\clearpage

\section{Introduction}
 
Current measurements from the Large Hadron Collider (LHC) are starting to put stringent bounds on the supersymmetry (SUSY) spectrum and cross sections. In particular, the bounds on the masses of scalar particles are reaching the $\TEV$ scale. For instance, ATLAS and CMS constraints from inclusive squark and gluino searches in scenarios with conserved R-Parity and neutralino LSP can be found in \cite{ATLAS-CONF-2013-047,Chatrchyan:2014lfa}, in which ATLAS  excludes squark masses up to 1.7 TeV in mSUGRA/CMSSM scenarios for $m_{\tilde{g}}=m_{\tilde{q}}$. Scenarios considering R-Parity violation (RpV) have been recently investigated in \cite{ATLAS-CONF-2013-091,Chatrchyan:2013xsw}, in which CMS excludes top squark masses below 1020(820) GeV depending on the specific RpV couplings. Also, long-lived squark and gluino R-hadrons have been studied in \cite{Aad:2013gva,Chatrchyan:2012sp}, in which CMS excludes top squark masses below 737(714) GeV depending on the selection criteria. Notice that gluino bounds are similar in magnitude and even harder than squark constraints. However, sleptons are in general less constrained (see, for instance, \cite{Fuks:2014nha,Heisig:2013rya} for a phenomenological analysis). In the chargino-neutralino (weakino) sector, the constraints from direct production are less stringent because production cross sections are smaller. For instance, ATLAS and CMS constraints on chargino and neutralino masses in R-Parity conserved models with gaugino LSP have been investigated in \cite{Aad:2014vma,ATLAS-CONF-2013-028,Aad:2014nua,ATLAS-CONF-2013-093,CMS-PAS-SUS-13-006,CMS-PAS-SUS-13-017} and  R-Parity violating models are studied in \cite{ATLAS:2012kr,Chatrchyan:2012mea}. Without loss of generality, weakinos which are much lighter than $1 \TEV$ are still allowed. 

Having this in mind, in this work we consider a nowadays empirically attractive flavor of SUSY, which is denominated Split Supersymmetry (Split-SUSY) \cite{ArkaniHamed:2004fb,Giudice:2004tc}. In this setup, the mass of scalars except for the Higgs boson are placed universally at the scale $\widetilde{m}$, which is high enough to account for collider bounds on squarks and sleptons. Concerning the gaugino sector, we require moderate weakino masses and a relatively heavier gluino. A detailed study of LHC constraints on Split-SUSY is beyond the scope of this work. Preliminary studies in this direction can be found in \cite{Jung:2013zya,Alves:2011ug,Wang:2006gp,Kilian:2004uj,Gupta:2004dz,Gambino:2005eh,Cheung:2004ad,Hewett:2004nw}. In Split-SUSY, since $\widetilde{m} \gg 1 \TEV$ the Higgs mass has to be fine-tuned. Nevertheless, in the original articles it is argued that there is a much larger fine-tuning associated to the cosmological constant. It is also worth recalling that this model retains interesting properties, such as gauge couplings unification, naturally suppressed flavor mixing, and a dark matter candidate. Furthermore, considering that the ATLAS and CMS collaborations have reported the observation of a Higgs-like particle \cite{Aad:2012tfa,Chatrchyan:2013lba}, it has been shown that in Split-SUSY it is possible to accommodate the observed Higgs mass \cite{Cottin:2012is,Giudice:2011cg,ArkaniHamed:2012gw,Arvanitaki:2012ps}, which imposes some constraints in the plane $(\widetilde{m},\tan\beta)$.

Concerning the neutrino sector, it has been shown that square mass differences and mixing angles can be reproduced in Split-SUSY by introducing BRpV terms plus a gravity-inspired operator \cite{Diaz:2009yz}. In this work we study this 
approach using updated data on neutrino observables \cite{Forero:2014bxa}. Consequently, the presence of these 
BRpV terms implies that the lightest Split-SUSY neutralino is overly unstable and therefore unable to play the role of 
dark matter. On the bright side, we show that the neutralino life time is short enough to avoid 
Big Bang Nucleosynthesis (BBN) restrictions \cite{Kawasaki:2004qu,Jedamzik:2006xz,Kawasaki:2008qe,Covi:2009bk}.

In order to account for the dark matter paradigm, we extend the Split-SUSY with BRpV model by adding a minimally 
coupled gravitino sector. We consider that gravitinos are thermally produced during the reheating period which follows 
the end of inflation, for which we mostly follow \cite{Rychkov:2007uq,Pradler:2007ne}. We show that, as long as $\widetilde{m}$ is small enough in comparison to the reheating temperature $T_R$, the standard expressions for the thermal gravitino relic density are still valid in our scenario. This is due to the fact that Split-SUSY is equivalent to the MSSM at energy scales greater than $\widetilde{m}$. Interestingly, this condition on $\widetilde{m}$ is quite consistent with the requirements obtained from computations of the Higgs mass. Let us note that there are other Split-SUSY models, where Higgsinos are also heavy \cite{ArkaniHamed:2012gw}, where it is possible to reconcile larger values of $\widetilde{m}$ with the Higgs mass. Thus, we use the standard expressions for the gravitino relic density, but consider the Split-SUSY RGEs \cite{Giudice:2004tc} for the parameters involved, in order to verify that this model reproduces the current values for the dark matter density \cite{Ade:2013zuv}. Moreover, taking into account the R-Parity violating gravitino-matter interactions we address the finite gravitino life time in detail. It becomes important to compute this quantity because we need a meta-stable dark matter candidate. It is also necessary to contrast the gravitino decay, together with its branching ratios to different final states, against experimental constraints coming from indirect dark matter detection via gamma rays, electron-positron pairs, neutrinos, etc. \cite{Grefe:2011dp,Choi:2010jt}. Since the gravitinos considered in our work are very long-lived, their potential effects in the early universe \cite{Khlopov:1984pf,Falomkin:1984eu,Khlopov:2004tn} are not further studied. 
 
The paper is organized as follows. In Section 2 we summarize the low energy Split-SUSY with BRpV setup together with the main features concerning the computation of the Higgs mass, neutrino observables, and the neutralino decay. In Section 3 we explain the details of the computation of the gravitino relic density and show that it is equivalent, up-to RGE flow of the parameters, to standard MSSM calculations. Moreover, we compute the gravitino life time in order to study the viability of the gravitino as a dark matter candidate, but also to show the particular interplay between neutrino physics and gravitino dark matter in our scenario. Conclusions are stated in Section 4.

\section{Split Supersymmetry}

Split Supersymmetry is a low-energy effective model derived from the MSSM, with all the
sfermions and all the Higgs bosons, except for one SM-like Higgs boson, decoupled at a scale 
$\widetilde m \gg 1\,\TEV$. The latest experimental results from the LHC favors this model, since no light sfermions have been seen in the laboratory. On the contrary, Split-SUSY makes no restriction on the masses of the charginos and neutralinos, and they can be as light as the electroweak scale. This is not contradicted by the 
experimental results, since direct production of neutralinos and charginos have a smaller 
production cross-section, and for this reason the constraints on their masses are less 
restrictive. 

The Split-SUSY Lagrangian is given by the following expression,

\begin{eqnarray}
{\cal L}^{split}_{susy}&=& {\cal L}^{split}_{kinetic}  +  m^2H^\dagger 
H - \frac{\lambda}{2}(H^\dagger H)^2 -\Big[ Y_u \overline q_L u_R i \sigma_2 H^*  +  Y_d \overline q_L d_R H  
+ \nonumber \\ 
&&
 Y_e \overline l_L e_R H  + \frac{M_3}{2} \widetilde G\widetilde G  +  \frac{M_2}{2} \widetilde W 
\widetilde W + \frac{M_1}{2} \widetilde B \widetilde B +  
\mu \widetilde H_u^T i \sigma_2 \widetilde H_d  + 
\label{eq:LagSplit} 
\\ 
&&
+\textstyle{\frac{1}{\sqrt{2}}} H^\dagger (\tilde g_u \sigma\widetilde W +  \tilde g'_u\widetilde B)\widetilde H_u
  +  \textstyle{\frac{1}{\sqrt{2}}} H^T i \sigma_2 (-\tilde g_d \sigma \widetilde W+ \tilde g'_d \widetilde B)\widetilde H_d 
+\mathrm{h.c.}\Big], \nonumber
\end{eqnarray}

\noindent where first we have the Higgs potential for the SM-like Higgs field $H$, with $m^2$ the Higgs mass
parameter and $\lambda$ the Higgs self interaction. Second, we have the Yukawa interactions, where 
$Y_u$, $Y_d$ and $Y_e$ are the $3\times3$ Yukawa matrices, which give mass to the up and down 
quarks and to the charged leptons after the Higgs field acquires a non-zero vacuum expectation value (vev).
As in the SM, this vev satisfies $\langle H \rangle=v/\sqrt{2}$, with $v^2=2m^2/\lambda$ and $v=246 \GEV$.
In the second line of Eq.~(\ref{eq:LagSplit}) we see the 3 gaugino mass terms and the Higgsino mass 
parameter $\mu$. Finally, in the third line we have the new couplings $\tilde g_u$, $\tilde g'_u$,
$\tilde g_d$ and $\tilde g'_d$ between Higgs, gauginos, and Higgsinos. These couplings are related
to the gauge couplings through the boundary conditions,

\begin{eqnarray}
\tilde g_u(\tilde m)=g(\tilde m) \sin\beta,
&\qquad&
\tilde g_d(\tilde m)=g(\tilde m) \cos\beta,
\nonumber\\
\tilde g'_u(\tilde m)=g'(\tilde m) \sin\beta,
&\qquad&
\tilde g'_d(\tilde m)=g'(\tilde m) \cos\beta,
\label{gtildeBC}
\end{eqnarray}

\noindent valid at the Split-SUSY scale $\widetilde m$. The couplings diverge from these boundary 
conditions due to the fact that the RGE are affected by the decoupling of the sfermions and heavy 
Higgs bosons. The angle $\beta$ is defined as usual by the relation $\tan\beta=v_u/v_d$, where 
$v_u/\sqrt{2}$ and $v_d/\sqrt{2}$ are the vevs of the two Higgs fields $H_u$ and $H_d$. Note that 
this definition makes sense only above the scale $\widetilde m$.

In the following sections we are going to study some observables that are useful for constraining the Split-SUSY scenario. However, as we are also interested in neutrino physics and cosmology, we are required to extend the previous Lagrangian by adding BRpV contributions and a gravitino sector. Therefore, for numerical results and figures we consider a scan on the 14-dimensional extended Split-SUSY parameter space, which contains the three soft masses, the Higgsino mass, the Split-SUSY scale, $\tan\beta$, six BRpV parameters and the reheating temperature. The details of this scan are shown in appendix~\ref{app:montecarlo}, where we specify the free parameters and some motivations for the chosen intervals. Here we just want to emphasize that some important experimental constraints, such as the Higgs mass, updated neutrino observables, and the dark matter density, are considered for the final selection of points. A systematic study of collider constraints considering the successful points of this scan is left for a future work. 

\subsection{Higgs mass}

We start this section pointing out that the latest ATLAS SM combined results report a value of $m_H = 126.0 \pm 0.4 (\text{stat}) \pm 0.4 (\text{sys}) \GEV$  \cite{Aad:2012tfa}. Meanwhile, CMS collaboration has reported $m_H = 125.3 \pm 0.4 (\text{stat}) \pm 0.5 (\text{syst}) \GEV$ \cite{Chatrchyan:2013lba}. Thus, in our scan we impose the constraint $125 \GEV < m_H < 127 \GEV$ to every point in the parameter space.

In order to properly compute the Higgs mass in Split-SUSY, we consider the effects of heavy decoupled particles on the theory at low energy, through the matching conditions at $\widetilde{m}$, plus the resum of large logarithmic corrections proportional to $\log(\widetilde{m}/M_{\text{weak}})$ by means of the renormalization group equations (RGEs). Thus, in practice, we follow the same procedure as \cite{Bernal:2007uv,Cottin:2012is}. Considering these aspects, we determine the scale-dependent Lagrangian parameters that are relevant for the Higgs mass computations. Afterwards, the Higgs mass is obtained from the sum of the tree-level contribution, which is proportional to $\lambda(Q)$, plus quantum corrections resulting from top and gaugino loops. Although the computed Higgs mass should be independent of the scale, it turns out that the truncation of the computations at some finite loop order produces a small scale dependence. Thus, it is customary to choose $Q=M_t$, with $M_{t}$ the top mass, because at this scale the quantum corrections are sub-leading.
 
We compute the quartic coupling $\lambda(M_t)$, and the rest of couplings and masses which are necessary to compute the radiative corrections to the Higgs mass, by using our own numerical implementation of Split-SUSY RGEs, which we take from reference \cite{Giudice:2004tc}. Besides the matching conditions for Split-SUSY gauge couplings given in Eq.~(\ref{gtildeBC}), we also take into account the matching condition for $\lambda(Q)$ at $\widetilde{m}$, 

\begin{equation}
\lambda(\widetilde m) = \frac{1}{4} \left[
g^2(\widetilde m) + g'^2(\widetilde m) \right] \cos^2 2\beta,
\label{ew:mhiggsinss}
\end{equation}

\noindent which relates the quartic Higgs coupling to the gauge couplings and $\tan\beta$. In Split-SUSY the threshold corrections to the Higgs quartic coupling resulting from integrating out the stops are very small, thus the boundary value for $\lambda(\widetilde m)$ is just given by the tree level value. Since the boundary conditions are given at different scales, we solve the RGEs by an iterative algorithm that finalizes when the numerical values of the considered parameters converge.

\begin{figure}[!ht*]
\begin{center}
\includegraphics[width=0.8\textwidth,angle=0]{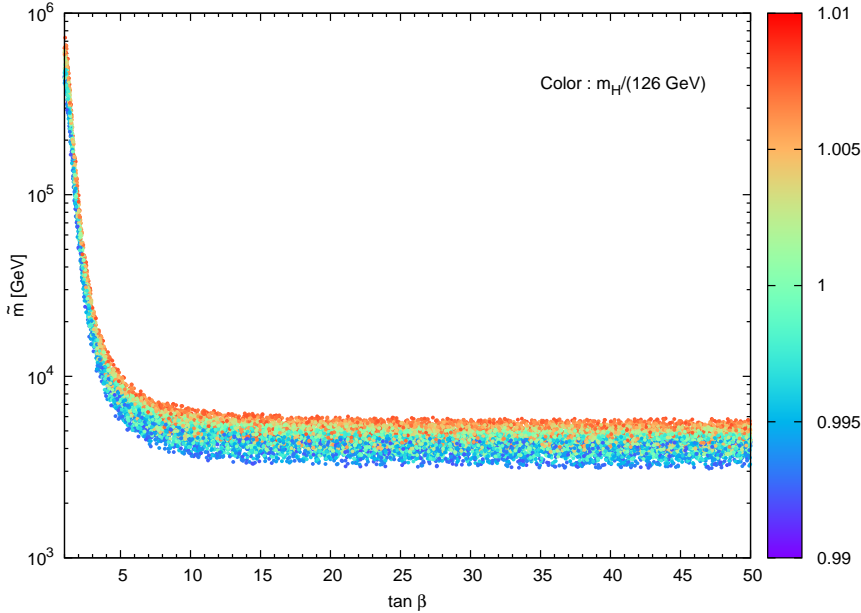}
\caption{Correlation between $\widetilde{m}$ and $\tan\beta$ in Split-SUSY derived from the observed Higgs mass constraint. In color code we plot the value of the ratio $m_H/126 \GEV$.}
\label{fig:MtildeTbeta}
\end{center}
\end{figure}

In Fig.~(\ref{fig:MtildeTbeta}) we check the correlation between $\widetilde{m}$ and $\tan \beta$ derived from the Higgs mass constraint in Split-SUSY, which for instance can be compared to \cite{ArkaniHamed:2004fb,Bernal:2007uv,Arvanitaki:2012ps,Djouadi:2013lra}. We see that the Split-SUSY scale approaches $4 \TEV$, as minimum, for increasing values of $\tan\beta$. Indeed, it must satisfy $\widetilde m \leq 8\,\TEV$ already for $\tan\beta \geq 8$. However, greater values of $\widetilde m$ can be obtained if $\tan\beta \leq 8$, with a maximum of $\widetilde{m} \simeq 10^6\,\TEV$. Note that small values for $\tan\beta$ are not ruled out in Split-SUSY as it is the case in the MSSM. The reason is that the squarks are not there to cancel out the quantum contribution from the quarks to the Higgs mass. Furthermore, constraints on $\tan\beta$ from heavy Higgs searches at the LHC or LEP, are not applicable to Split-SUSY.

\subsection{Neutrino masses and mixings}
\label{sec:neutrinos}

We are interested in generating neutrino masses and mixing angles, thus we include in the model described by Eq.~(\ref{eq:LagSplit}) bilinear R-Parity violating (BRpV) terms \cite{Hirsch:2000ef,Diaz:2006ee}. Notice that BRpV terms are also useful in order to relax cosmological constraints on the axion sector \cite{Hasenkamp:2011xh}. The relevant terms in the Lagrangian are
\begin{equation}
{\cal L}_{SS}^{RpV} \owns
\epsilon_i\widetilde H^T_u i\sigma_2 L_i
-\frac{1}{\sqrt{2}} a_i H^T i \sigma_2
(-\tilde g_d \sigma\widetilde W+\tilde g'_d\widetilde B)L_i \ + \ \mathrm{h.c.},
\label{LSplitRpV}
\end{equation}
where we have the three BRpV parameters $\epsilon_i$, which have units of mass. The dimensionless
$a_i$ parameters are equivalent to the sneutrino vacuum expectation values, and they appear 
in the Split-SUSY with BRpV model after integrating out the heavy scalars. Their existence implies
that we are assuming that the low energy Higgs field $H$ has a small component of sneutrino \cite{Cadiz:2013yja}. The corresponding matching condition is given by

\begin{equation}
  a_{i}(\widetilde{m}) = \frac{s_i}{\cos\beta},
\label{eq:mca}
\end{equation}

\noindent where $s_{i} = v_{i}(\widetilde{m})/v$ with $v_{i}$ the vacuum expectation value of sneutrinos. 

Neutrinos acquire mass through a low energy see-saw mechanism, where neutrinos mix with the neutralinos in a $7\times7$ mass matrix,
\begin{equation}
{\cal M}_N=\left(\begin{array}{cc} {\mathrm M}_{\chi^0} & m^T \\ 
m & 0 \end{array}\right),
\label{X07x7}
\end{equation}
written in the basis $(-i\widetilde B,i\widetilde W^0,\widetilde H_d^0, \widetilde H_u^0,
\nu_e, \nu_\mu, \nu_\tau)$. The high energy scale is given by the neutralino masses, described by the 
$4\times4$ mass matrix,
\begin{equation}
{\mathrm M}_{\chi^0}=\left(\begin{array}{cccc}
M_1 & 0 & -\frac{1}{2}\tilde g'_d v & \frac{1}{2}\tilde g'_u v \\
0 & M_2 & \frac{1}{2}\tilde g_d v & -\frac{1}{2}\tilde g_u v \\
-\frac{1}{2}\tilde g'_d v & \frac{1}{2}\tilde g_d v & 0 & -\mu \\
\frac{1}{2}\tilde g'_u v & -\frac{1}{2}\tilde g_u v & -\mu & 0
\end{array}\right),
\label{X0massmat}
\end{equation}
which is analogous to the neutralino mass matrix of the MSSM, with the main difference in the 
D-terms: since there is only one low-energy Higgs field, these terms are proportional to its vev $v$.
In addition, these terms are proportional to the Higgs-gaugino-Higgsino couplings described earlier.
The BRpV terms included in the Split-SUSY Lagrangian generate the mixing matrix $m$,
\begin{equation}
m=\left(\begin{array}{cccc}
-\frac{1}{2} \tilde g'_d a_1v & \frac{1}{2} \tilde g_d a_1v 
& 0 &\epsilon_1 \cr
-\frac{1}{2} \tilde g'_d a_2v & \frac{1}{2} \tilde g_d a_2v&0 
& \epsilon_2 \cr
-\frac{1}{2} \tilde g'_d a_3v & \frac{1}{2} \tilde g_d a_3v&0 
& \epsilon_3
\end{array}\right).
\label{X0mixingm}
\end{equation}
We see here the supersymmetric parameters $\epsilon_i$ and the effective 
parameters $a_i$. It is well known that this mixing leads to an effective $3\times3$ neutrino mass matrix
of the form
\begin{equation}
{\bf M}_\nu^{eff}=-m\,{\mathrm{M}}_{\chi^0}^{-1}\,m^T=
\frac{v^2}{4\det{{\mathrm M}_{\chi^0}}}
\left(M_1 \tilde g^2_d + M_2 \tilde g'^2_d \right)
\left(\begin{array}{cccc}
\lambda_1^2        & \lambda_1\lambda_2 & \lambda_1\lambda_3 \cr
\lambda_2\lambda_1 & \lambda_2^2        & \lambda_2\lambda_3 \cr
\lambda_3\lambda_1 & \lambda_3\lambda_2 & \lambda_3^2
\end{array}\right),
\label{treenumass}
\end{equation}
where we have defined the parameters $\lambda_i= a_i\mu+\epsilon_i$. This mass matrix has only one non-zero
eigenvalue, and thus only an atmospheric mass scale is generated. 

In Split-SUSY with BRpV, even if we add one-loop quantum corrections, the solar mass difference remains equal to zero \cite{Diaz:2006ee}. The quantum corrections only modify the atmospheric mass difference, that is already generated at tree level. On the contrary, in the MSSM with BRpV, we can generate a solar neutrino mass difference through quantum corrections \cite{Hirsch:2000ef}. Finally, we mention that in models with spontaneous violation of R-Parity, we can generate both atmospheric and solar mass differences at tree level, see for instance \cite{Hayashi:1984rd,Masiero:1990uj,Romao:1992vu,Romao:1991ex,Kitano:1999qb,Vicente:2009wk,Barger:2010iv,Marshall:2014kea} and references therein. In these models the $\epsilon_i$ are generated dinamically thanks to the introduction of extra fields that also make the neutrino effective mass richer.

\begin{figure}[!ht*]
\begin{center}
\includegraphics[width=0.8\textwidth,angle=0]{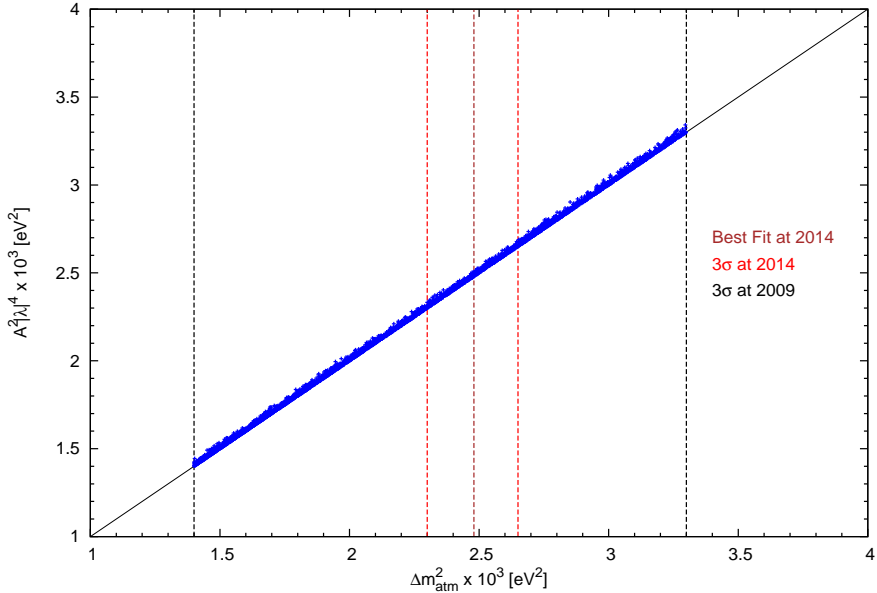}
\caption{Atmospheric mass correlation. The output of our scan corresponds to the value of $\Delta m^2_{\text{atm}}$, which is shown along the x-axis just to simplify the reading of experimental $3\sigma$ bounds.}
\label{fig:AtLambda_ug}
\end{center}
\end{figure}

In Split-SUSY with BRpV, a possible way to account for a tree level solar mass is the introduction of a dimension-5 operator that gives a mass term to the neutrinos after the Higgs field acquires a vev \cite{Vissani:2003aj,Diaz:2009yz,Cottin:2011fy}. This operator may come from an unknown quantum theory of gravity. Assuming a lower than usual Planck scale, as suggested by theories with extra dimensions, the contribution to the neutrino mass matrix can be parametrized as follows 
\begin{equation}\label{muMatrix}
\Delta M_g^\nu = \mu_g
\left[\begin{array}{cccc}
1 & 1 & 1 \cr
1 & 1 & 1 \cr
1 & 1 & 1 
\end{array}\right],
\end{equation}
where $\mu_g\simeq v^2/\bar{M}_{P}$, with $\bar{M}_{P}$ the effective Planck scale, gives the overall scale of the contribution, and flavor-blindness from gravity is represented by the matrix where all the entries are equal to unity. This scenario is not modified by the block diagonalization
of neutralinos and neutrinos in the BRpV scenario. Thus, we are formally assuming that this contribution is democratic in 
the gauge basis. Then, the effective neutrino mass matrix is given by,

\begin{equation}
\label{eq_mixmod}
M_\nu^{ij}=A \lambda^i\lambda^j+\mu_g.
\end{equation}
In this work we re-compute the values for the parameters $A$, $\lambda_{i}$ and $\mu_{g}$ in order to account for up-to-date measurements of neutrino observables \cite{Forero:2014bxa}. First of all, we notice that successful points for neutrino
observables considered in reference \cite{Diaz:2009yz} are not necessarily consistent with updated $95\%$ confidence level intervals. Nonetheless, we are still able to find several points with this confidence level that are also consistent with the analytical expressions for square mass differences and mixing angles derived in this reference. These expressions are given by,

\begin{eqnarray}
  \Delta m_{\text{atm}}^2 &\simeq& A^2 |\lambda|^4, \nonumber \\
  \Delta m_{\text{sol}}^2 &\simeq& \mu_{g}^2\frac{(\vec{v}\times\vec{\lambda})^4}{\vec{\lambda}^4}, \nonumber \\ 
  \sin^2 \theta_{\text{reac}} &\simeq& \lambda_{1}^2/|\vec{\lambda}|^2,    \label{eq:neuobs} \\
  \tan^2 \theta_{\text{atm}}  &\simeq& \lambda_{3}^2/\lambda_{2}^2, \nonumber \\
  \tan^2 \theta_{\text{sol}}  &\simeq& \frac{\lambda_{2}^2+\lambda_{3}^2}{(\lambda_{3}-\lambda_{2})^2}, \nonumber 
\end{eqnarray}

\noindent which are valid approximations in the regime $A|\lambda|^2 \gg \mu_g$ and $\lambda_{1}^2 \ll \lambda_{2}^2 + \lambda_{3}^2$, with $\vec{v}=(1,1,1)$. Note that this regime is naturally selected by a blind Monte Carlo scan. For instance, in Fig.~(6) of \cite{Diaz:2009yz} it can be seen that the solar mixing angle is the main reason for selecting this zone of the parameter space. In our scan, we use these approximations to speed-up the search of successful points because we think that this approximation is quite unique considering the neutrino mass matrix given in Eq.~(\ref{eq_mixmod}). Interestingly, the condition $A|\lambda|^2 \gg \mu_g$ also implies that only normal hierarchy is allowed in our model. In Fig.~(\ref{fig:AtLambda_ug}) the correlation between the combination $A^2 |\lambda|^4$ and the atmospheric mass is shown, where we highlight the effects of observational improvements on the allowed intervals for the input parameters.

\subsection{Unstable neutralino}

The introduction of BRpV terms in the Split-SUSY Lagrangian, which are useful in order to explain neutrino physics, implies that the lightest SUSY particle is unstable. Therefore, the usual approach of considering the neutralino as a dark matter candidate is essentially forbidden in this scenario. In order to illustrate this point, we explicitly calculate the life time of the lightest neutralino. In practice, it is enough to consider on-shell, two-body decay channels, which are given by $\chi^0_{1}\rightarrow H\nu,\,Z\nu,\,W^{\pm}l^{\mp}$. The corresponding Feynman rules are given in appendix \ref{app:neutralinodecay}. It is worth noting that the neutralino-neutrino couplings in the neutrino mass basis involve, in general, the $U_\text{PMNS}$ mixing matrix. However, we do not consider this dependence because it disappears when we sum over all neutrino species, since $U_\text{PMNS}$ is unitary. Also, it is interesting to note that all the couplings are independent of $\epsilon_i$ (see details in appendix \ref{app:neutralinodecay}). We notice that the contributions of R-Parity conserving decay channels, involving the decay of the neutralino into gravitino and SM particles, are much smaller. Indeed, effective life times associated to these kind of channels are typically above $1$~sec \cite{Covi:2009bk}, while those associated to R-Parity violating channels are smaller than $10^{-11}$~sec. Therefore, the correlations between neutralino BRs and neutrino mixing angles remain valid (at least at tree level). 

\begin{figure}[!ht*]
\begin{center}
\includegraphics[width=0.8\textwidth,angle=0]{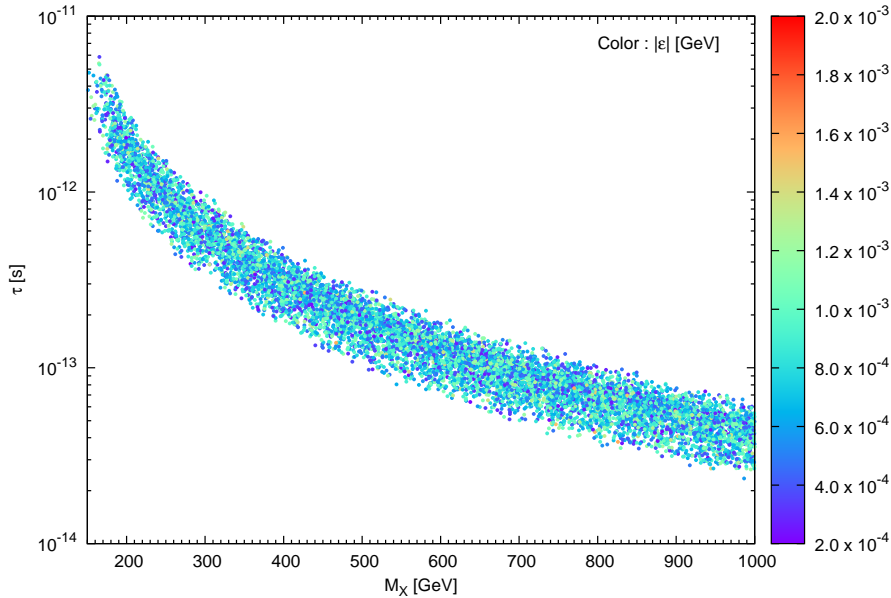}
\caption{Neutralino life time as a function of its mass in Split-SUSY with BRpV. In color we show the values that span the parameters $\epsilon_i$. It can be seen that there is no correlation between the values of the neutralino life time and these parameters.}
\label{fig:tauNmassN}
\end{center}
\end{figure}

In Fig.~(\ref{fig:tauNmassN}) we show the results concerning the neutralino life time obtained from our scan. We see that the life time{is shorter than $10^{-11}\,\text{sec}$ in the whole mass range. This result is sufficient to verify that a neutralino dark matter candidate, with $m_{\chi}\geq 200\GEV$, is not allowed in Split-SUSY with BRpV because it decays overly fast. On the other hand, we are quietly safe from BBN constraints on unstable neutral particles \cite{Kawasaki:2004qu,Jedamzik:2006xz,Kawasaki:2008qe,Covi:2009bk}, which are easily avoided when the life time is shorter than $0.01\,\text{sec}$. Considering that BBN constraints are not an issue, in the following section we introduce a gravitino as the lightest SUSY particle (LSP) in order to account for dark matter. 

However, before moving to the gravitino dark matter section, let us briefly comment about the phenomenology of short-lived neutralinos in the context of collider searches at the LHC. When the neutralino life time is larger than $10^{-12}\,\text{sec}$ ($m_{\chi^0} \leq 300\GEV$ in our scenario) the produced neutralinos should decay inside the inner detector of ATLAS \cite{ATLAS-CONF-2013-092}, allowing the reconstruction of the corresponding displaced vertex (CMS requires $\tau\geq10^{-11}\,\text{sec}$ \cite{CMS-PAS-EXO-12-037}). Instead, if the life time is much shorter, the collider searches must rely on the measurement of some excess of events over SM background in multi-lepton or multi-jet channels. In order to interpret these searches in our scenario, it is useful to know the neutralino BRs. In our scan we assume that the decay width is mostly accounted for by the on-shell final states $Z\nu$, $H\nu$ and $W^{\pm}l^{\mp}$. Then, the corresponding BRs are computed with respect to the sum of these three channels. The results are shown in Fig.~(\ref{fig:BRmassN}). It can be seen that light neutralinos prefer to decay into $Z\nu$, especially in the region of displaced vertex searches (notice that the neutrino makes this search quite challenging). However, when the neutralino is heavy enough, the BRs into $Z\nu$ and $H\nu$ become very similar and close to $40\%$ each. The BR into $W^{\pm}l^{\mp}$ is in general sub-dominant and rarely exceeds the $30\%$ level. These results can be used as guidance for the search of neutralinos at the LHC, but should be considered in combination with the life time and production cross sections to constrain the model.

\begin{figure}[!ht*]
\begin{center}
\includegraphics[width=0.8\textwidth,angle=0]{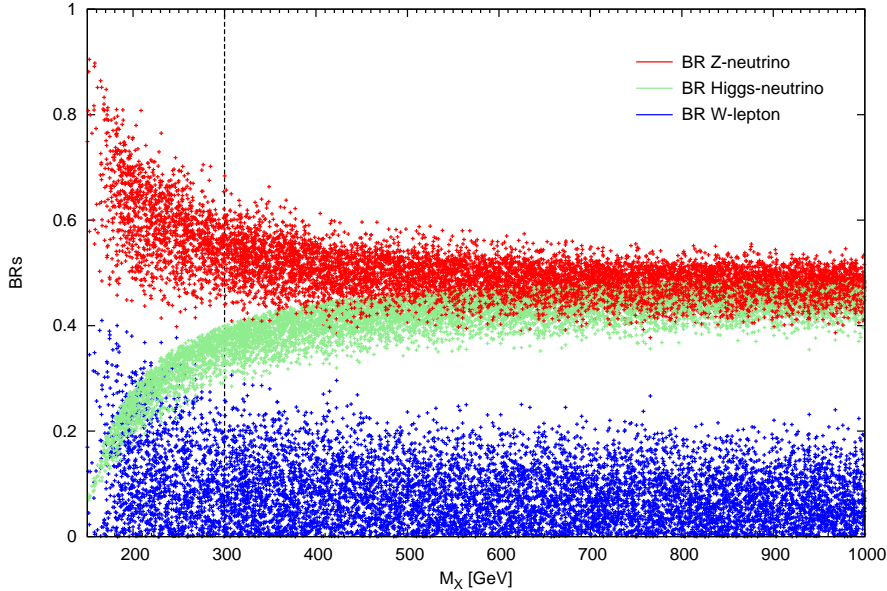}
\caption{Neutralino branching ratios. To the left side of the vertical dotted line the neutralino life time is able to be larger than $10^{-12}\,\text{sec}$, allowing collider searches based on the reconstruction of a displaced vertex. To the right side, the life time is much shorter and the neutralino would decay almost instantaneously after being produced.}
\label{fig:BRmassN}
\end{center}
\end{figure}

\section{Gravitino Cosmology}

Supersymmetric models with a gravitino-LSP represent an attractive scenario in which to accommodate dark matter observations \cite{Fayet:1981sq,Giudice:1999am,Bolz:2000fu,Pradler:2006qh,Heisig:2013sva,Takayama:2000uz,Buchmuller:2007ui,Diaz:2011pc,Restrepo:2011rj}. For instance, the thermal gravitino relic density has been computed and successfully connected to observations in the context of conserved R-Parity \cite{Giudice:1999am,Bolz:2000fu,Pradler:2006qh,Heisig:2013sva}. Furthermore, it has been shown that even allowing R-Parity violating terms these results still remain positive \cite{Takayama:2000uz,Buchmuller:2007ui,Diaz:2011pc,Restrepo:2011rj}.  Although in RpV scenarios the gravitino is allowed to decay at low temperatures, it has been shown that in general it remains stable for timescales comparable to the age of the universe. Indeed, the gravitino is naturally meta-stable since the interactions are doubly suppressed by both the smallness of the RpV terms and the Planck mass. 

In the context of Split-SUSY, we consider that in order to compute the thermal gravitino relic density it is necessary to study the potential effects of the mass scale $\widetilde{m} \gg 1\,\TEV$ for scalar sparticles. For instance, the introduction of this arbitrary scale implies that during the thermal history of the universe there is an abrupt change in the number of relativistic degrees of freedom around the temperature $T\simeq \widetilde{m}$. This behavior is remarkably different to the natural MSSM scenarios\footnote{We assume that in natural MSSM scenarios the supersymmetric scale is common and fixed at the $\TEV$ scale. Thus, every supersymmetric particle have a mass around the $\TEV$ value.} and in principle it may affect the standard computations of the gravitino relic density. In this section we show that under reasonable assumptions, concerning the Split-SUSY scale and the reheating temperature, the relic density formula in Split-SUSY is equivalent to the MSSM result. In addition, we compute the gravitino life time in order to study the particular interplay between neutrino physics and dark matter obtained in our scenario. 

\subsection{Gravitino relic abundance}
\label{sec:relicabundance}

We assume that the evolution of the early universe is determined by the standard model of cosmology \cite{Kolb:1990vq,steven2008cosmology}. In this approach, it is assumed that after the Big Bang the universe experiences an inflationary phase, which is triggered by the dynamics of a slow-rolling scalar field \cite{PhysRevD.23.347,Linde:1981mu,PhysRevLett.48.1220}. During this phase, the energy density is dominated by vacuum energy, such that the universe expands exponentially. Also, it is commonly assumed that any trace of pre-inflationary matter and radiation is diluted to negligible levels with the corresponding supercooling of the universe. The inflationary phase concludes when the inflaton field reaches the bottom of the scalar potential, such that the universe becomes matter dominated. During this stage, the inflaton starts to decay more rapidly into other forms of matter and radiation, giving rise to the radiation dominated phase.

In general, it is possible to conceive that during the last stages of inflation, some amount of gravitinos are non-thermally generated through inflaton decays. Moreover, it has been shown that this mechanism can be very effective \cite{Kallosh:1999jj,Giudice:1999am,Asaka:2006bv}, and even dangerous concerning observational constraints. However, these results are strongly dependent on the inflationary model. For example, it is possible to construct a supersymmetric model where this mechanism of production can be safely neglected \cite{Nakayama:2012hy}. Thus, in order to avoid deeper discussions about the inflationary model, this mechanism of production is not considered in the realm of this work. 

Leaving aside the non-thermal production of gravitinos, it turns out that the total amount of energy stored in the inflaton field is progressively transformed into relativistic matter. This process increases dramatically the temperature and entropy of the universe. When this phase is completed, the universe reaches the reheating temperature $T_R$. Depending on the magnitude of $T_R$, the gravitinos could arrive at thermal equilibrium with their environment during the post-reheating period. In this case, a very light gravitino may account for the observed DM relic density. However, it has been shown that this scenario is quite difficult to achieve \cite{Staub:2009ww}. Instead, we assume that the gravitino is out of thermal equilibrium and, in addition its initial number density is required to be negligible. Therefore, the gravitino relic density is generated from the scattering and decays of particles, which are indeed in thermal equilibrium in the plasma. 

In order to compute the gravitino relic density we consider the approach of \cite{Rychkov:2007uq}\footnote{In the first stages of our work we have considered \cite{Pradler:2006qh} and just recently we have been brought to \cite{Rychkov:2007uq}. The latter reference improves importantly the computations of the relic density in the regime $g_s(T_R) \sim 1$, which allows us to consider points with $T_R \sim 10^5 \GEV$.}. This reference considers a minimal version of supergravity in four dimensions with N=1 supersymmetry, or equivalently the MSSM with minimal gravitino interactions. Also, it is important to note that at the computational level it is used that $m_{\text{SUSY}} \ll T_R$, where $m_{\text{SUSY}}$ is the common mass scale of supersymmetric particles.  

For the following discussion it is convenient to consider an instantaneous reheating period\footnote{The main idea of the discussion is not modified by pre-reheating features because we are interested in temperatures $T'$ such that $T_R \gg T' \gg m_{\text{SUSY}}$. However, as the correction is approximately $25\%$ after considering non-instantaneous reheating, we use the full result for the relic density.}. Thus, the energy stored in the inflaton modes is suddenly transformed into radiation energy. This is equivalent to starting the thermal history of the universe from a Big Bang with a maximum temperature $T_R$. Then, the expression for the gravitino comoving density $Y_{3/2}(T)$, evaluated at a temperature $T'$ such that $T_R \gg T' \gg m_{\text{SUSY}}$, is given by

\begin{equation}
  Y_{3/2}(T') = \frac{n_{3/2}(T')}{s(T')} \hspace{1mm} = \hspace{1mm} -\int_{T_R}^{T'}dT\frac{C_{3/2}(T)}{s(T)H(T)T},
\end{equation}

\noindent where $n_{3/2}(T)$ is the gravitino number density, $s(T)\propto T^3$ is the entropy, $H(T)\propto T^2$ is the Hubble parameter and $C_{3/2}(T)$ is the collision factor that determines the rate of gravitino production at a given temperature. For the computation of this rate it is sufficient to consider interactions only between relativistic particles. This is because the number density of non-relativistic particles is exponentially suppressed and therefore their contributions can be neglected. Indeed, it is assumed and finally verified that the relevant temperatures for the computation of the relic density are much greater than $m_{\text{SUSY}}$, then the relativistic approximation is applied to every particle of the MSSM.

After pages of computations, considering several contributions to the collision factor, it is finally obtained that $C_{3/2}(T)\propto T^6$. Then $Y_{3/2}(T') = F(T_R)(T_R-T')$, where $F(T)$ is a function that varies very slowly with the temperature. Interestingly, this result indicates that the gravitino production is only efficient around the reheating epoch. In practice, it is enough to consider $T'$ sufficiently smaller than $T_R$ in order to obtain, with a good level of accuracy, the gravitino comoving relic density that is valid for any temperature in the future.

Therefore, considering that Split-SUSY is equivalent to the MSSM at scales greater than $\widetilde{m}$, we notice that within the region of the parameter space restricted by $\widetilde{m} \ll T_{R}$, the expression for the gravitino relic density in Split-SUSY is equivalent to the MSSM result. Thus, the normalized relic density is given by,

\begin{eqnarray}
  \Omega_{3/2}(T_{0}) h^2 &=& m_{3/2}Y_{3/2}(T_0)\frac{s(T_0)h^2}{\rho_c(T_0)} \label{eq:relicdensity} \\
  &=& 0.167 \bigg(\frac{m_{3/2}}{100\GEV}\bigg)\bigg(\frac{T_{\text{R}}}{10^{10}\GEV}\bigg)\bigg[\frac{1.30}{2\pi^5}
      9\lambda_t^2(T_R) +  \nonumber \\
  & &\sum_{N=1}^3\bigg(1+ \frac{M_{N}^{2}(T_R)}{3m_{3/2}^2}\bigg)\bigg(\frac{n_Nf_N(\alpha_Ng_N(T_R))}{2(2\pi)^3}+\frac{1.29}{8\pi^5}g_N^2(T_R)(C_N'-C_N)\bigg)\bigg], \nonumber 
\end{eqnarray}

\noindent where $T_0=2.725 K$ is the CMB temperature today, $s(T_0)=2.22\times10^{-38}\GEV^3$ is the current entropy density and $\rho_c(T_0)h^{-2}=8.096\times 10^{-47}\GEV^4$ is the critical density. The sum over $N$ takes into account the contribution from the degrees of freedom associated to the gauge groups of the MSSM, i.e. $U(1)$, $SU(2)$ and $SU(3)$ respectively. The functions $f_N(\alpha_Ng_N(T_R))$, with $\alpha_N=\{\sqrt{11}/2,3/2,3/2\}$, are the ``improved'' rate functions of the MSSM, which we get from Fig.~(1) of \cite{Rychkov:2007uq}. The other parameters are $n_N=\{1,3,8\}$, $C_N=\{0,6,24\}$ and $C_N'=\{11,21,48\}$. Also, we set $A_t(T_R)=0$. Notice that the gaugino masses $M_{N}$ and gauge couplings $g_{N}$ are evaluated at the reheating temperature.  

The complementary region of the parameter space, $\widetilde{m} > T_{R}$, is not further developed in our work. Indeed, this region is highly disfavored by taking into account the typical values of $T_R$ that are considered in models of baryogenesis through high scale leptogenesis \cite{Fukugita:1986hr,Buchmuller:2004nz}, and the values obtained for $\widetilde{m}$ from the Higgs mass requirement.

In our scan we generate every parameter of the model at low energy scales. For instance, gauge couplings are defined at $M_W$, the top Yukawa coupling at $M_t$ and soft masses at $M_{\text{ino}}$. Thus, in order to evaluate the couplings and gaugino masses at $T_R$, we use a three-step approach for the running of the parameters. We use the SM RGEs between $M_W$ and $M_{\text{ino}}$, then the Split-SUSY RGEs between $M_{\text{ino}}$ and $\widetilde{m}$ and finally the MSSM RGEs from $\widetilde{m}$ until $T_R$. In order to have control of the energy scales, we take the reheating temperature as a free parameter, then we assume that the gravitino saturates the dark matter density in order to obtain its mass from Eq.~(\ref{eq:relicdensity}). We have checked that by using Eq.~(\ref{eq:relicdensity}) instead of the expression given in \cite{Pradler:2006qh} we obtain gravitino masses which are as much as $50\%$ bigger. 

In Fig.~(\ref{fig:TrMassG}) we show the allowed relations between the reheating temperature, gravitino mass and gaugino scale in order to obtain the observed dark matter density, $\Omega_{DM} h^2= 0.1196 \pm 0.0031$ \cite{Ade:2013zuv}. We see that the boundary values of $m_{3/2}$ are positively correlated to the values of $M_{\text{ino}}$ for each value of the reheating temperature. From the Higgs mass requirement we have obtained that the minimum accepted Split-SUSY scale is $\widetilde{m}=4\times10^3\GEV$. Then, in order to satisfy the approximation $T_R \gg \widetilde{m}$, we consider reheating temperatures in the interval $10^2\widetilde{m} < T_{R}<10^6\widetilde{m}$. Consequently the minimum reheating temperature that we accept is $T_R^{\text{min}}=4 \times 10^5 \GEV$. At this temperature the strong coupling constant is approximately $g_s(T_R^{\text{min}}) \simeq 0.94$.

\begin{figure}[!ht*]
\begin{center}
\includegraphics[width=0.8\textwidth,angle=0]{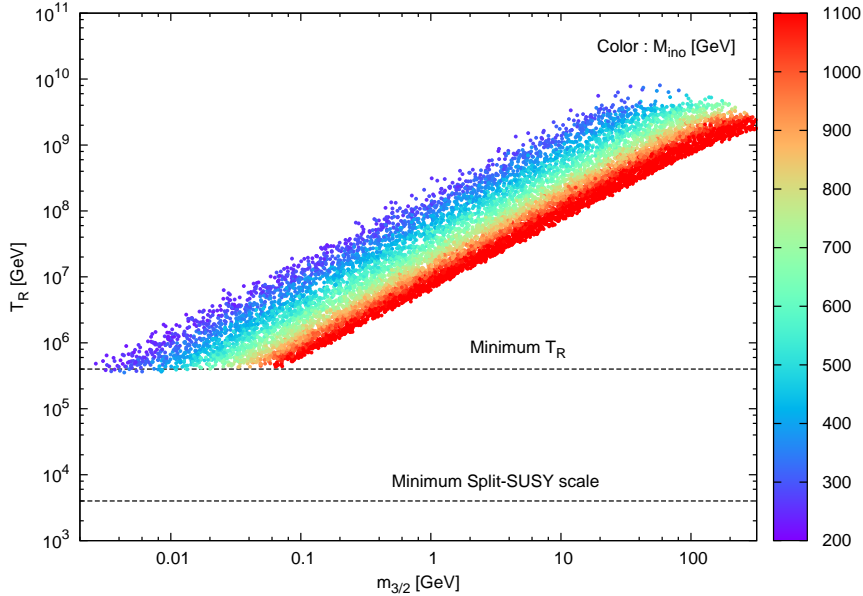}
\caption{Relation between the reheating temperature, gravitino mass and gaugino scale in order to obtain the observed dark matter density. The black dashed line at about $4 \TEV$ indicates the minimum value of $\widetilde{m}$ obtained from the Higgs mass requirement. In the Monte Carlo scan we assume $T_{R} \geq 10^2 \widetilde{m}$, then the lower bound on $T_{R}$ is around $10^{6} \GEV$, which is indicated by a second dashed line in the figure.}
\label{fig:TrMassG}
\end{center}
\end{figure}

Concerning the recent results from BICEP2 \cite{Ade:2014xna}, which suggest that the inflation energy scale is such that $V_{\text{inf}}^{1/4}(\phi) \simeq 2.2\times10^{16} \GEV$ for a tensor-to-scalar ratio $r\simeq0.2$, we must point out that the reheating temperature is not necessarily required to be that high. In generic inflationary models, the reheating temperature is mostly defined in terms of the inflaton decay width, which could be related to the inflaton potential but it depends essentially on the specific model of interactions between the inflaton and the rest of the particles. In our work we consider the inflaton decay width, or equivalently the reheating temperature, as a free parameter.

\subsection{Gravitino life time}

Once we allow BRpV terms in the Split-SUSY Lagrangian, every supersymmetric particle, including the gravitino, becomes unstable. Fortunately, it turns out that in the considered mass range, $m_{3/2} < m_H$, the gravitino life time is considerably larger than the age of the universe, which guarantees the necessary meta-stability required for the dark matter particle. However, the meta-stability of the gravitino opens the possibility for the indirect observation of dark matter through the detection of gamma-rays or charged particles of cosmic origin. Instead, the non-observation of any excess with respect to the background in these searches is useful for constraining the parameter space allowed by the model. Therefore, in order to study the experimental potential and consistency of our dark matter scenario it is unavoidable to consider the life time of the gravitino. 

Some studies concerning the gravitino life time and its experimental potential in the MSSM with RpV terms are given by \cite{Grefe:2011dp,Choi:2010jt,Buchmuller:2007ui,Restrepo:2011rj,Moreau:2001sr}. Moreover, the gravitino life time has been studied in the context of Partial Split-SUSY with BRpV \cite{Diaz:2011pc} considering all the available channels in the mass range $m_{3/2} < m_H$. Because of the similarity between the latter model and our scenario, we consider the nomenclature, mass range, and several intermediate results from this reference. However, there are explicit model-dependent features that deserve some consideration, such as the explicit form of the gravitino-to-matter couplings and the relation between BRpV parameters and neutrino observables. 

For the sake of clarity, the details about the computations of the gravitino decay in Split-SUSY with BRpV are reserved for appendix \ref{app:gravitinodecay}. Indeed, in order to study the main features of our scenario, it is enough to consider the factorized expression for the gravitino decay width, 

\begin{equation}
  \Gamma_{3/2} = \sum_{i \in U} g_i(m_{3/2})h_i(M_{1},M_{2},\mu,\tilde g_d,\tilde g'_d,|\lambda|^2),
\label{eq:gravitino_decay}
\end{equation}

\noindent where the index $i$ runs over all available channels in the range $m_{3/2} < m_H$, i.e. $U= \{\gamma\nu_j,W^*\nu_j,\gamma (Z^*)l_j\}$, where $j$ is a family index, and the two-body notation $W^*\nu_j$ or $\gamma(Z^*)l_j$ implies the sum over all three-body decays for a fixed $j$. The functions $g_i(m_{3/2})$ involve complicated integrals over the phase space of the three-body final states that in general must be evaluated numerically. However, each function $h_i$ is analytical and relatively simple. Furthermore, it is obtained that every factor $h_i$ is proportional to $|\lambda|^2$.

Thus we have that the total gravitino decay width is proportional to $|\lambda|^2$, as is is the general case in SUSY with BRpV scenarios. Interestingly, in Split-SUSY with BRpV the term $|\lambda|^2$ is proportional to the neutrino atmospheric mass but it is independent of the solar mass, check the first lines of Eq.~(\ref{eq:neuobs}). This is not the case in the MSSM or Partial Split-SUSY with BRpV, where the quantity $|\lambda|^2$ (or $|\Lambda|^2$ to be exact) can be related to the solar mass as well. As the life time is inversely proportional to the decay width and $\Delta m_{\text{atm}}^{2} \gg \Delta m_{\text{sol}}^{2}$, we expect typical values of the gravitino life time in Split-SUSY with BRpV to be smaller than the values obtained in the MSSM or Partial Split-SUSY with BRpV. Furthermore, we can use Eq.~(\ref{eq:neuobs}) and the approximation $\tilde{g} \ll 1$ in order to obtain

\begin{equation}
  \Gamma_{3/2} \simeq |\sqrt{\Delta m_{\text{atm}}^2}|\sum_{i \in U} g_i(m_{3/2})\bar h_i(M_{1},M_{2},\tilde g_d,\tilde g'_d),
\label{eq:gravitino_decay_firstapp}
\end{equation}

\noindent where the parameters $|\lambda|^2$ and $\mu$ were traded for the atmospheric mass, whose value is tightly restricted by neutrino experiments. This expression is useful for visualizing the subset of Split-SUSY with BRpV parameters that determine the gravitino life time. At first sight, we just need to fix the parameters ($m_{3/2}$, $M_1$, $M_2$, $\tilde g_d$, $\tilde g'_d$) in order to obtain both the life time and the corresponding BRs. Below, we test this hypothesis against exact numerical computations of $\Gamma_{3/2}$ and derive the effective set of parameters that do the job. Also, we include some approximated life time bounds for the considered scenarios in order to verify that at least light gravitinos are still allowed in our model.
 
Assuming that Eq.~(\ref{eq:gravitino_decay_firstapp}) is a valid approximation, we see that the gravitino branching ratios and the total decay width should be determined by $M_1$ and $M_2$ plus some noise coming from the values of $\tilde g_d$ and $\tilde g'_d$, which depend on $\tilde m$ and $\tan \beta$. Indeed, this feature can be checked in Fig.~(\ref{fig:tauG_BR3B_massG_M1_300}) (top panel), where we have fixed $M_1=300\,\GEV$ and $M_2=2M_1$ while we vary the rest of parameters as usual, and require the Higgs mass and neutrino physics to be satisfied. In this figure, we also show the bounds on the gravitino life time obtained in \cite{Choi:2010jt} for the same values of $M_1$ and $M_2$. These bounds are approximate because the previously cited reference only considers RpV terms in the tau sector. Below $10\GEV$, we consider the bounds obtained in \cite{Albert:2014hwa} in the regime $BR(\psi_{3/2} \rightarrow \gamma\nu)=1$. In order to construct the boundary line, we correct the life time bounds using our BRs. In practice, we just consider the BR-corrected values of $\tau^{\text{max}}_{3/2}$ for $m_{3/2}=1,5$ and $10\,\GEV$.

Extending the discussion to scenarios where $M_2 \neq 2M_1$, we have found that in general the distribution of points in the plane ($m_{3/2}$, $\tau_{3/2}$) is more diffuse, in particular when $M_2$ is very similar to $M_1$, because the noise from the couplings becomes more important. When this is the case we just need to fix $\tan\beta$ because $\widetilde{m}$ should be restricted by the Higgs mass value. An example of this kind of scenario can be seen in Fig.~(\ref{fig:tauG_BR3B_massG_M1_300}) (bottom panel). We see that the life time and three-body BR are fixed for each value of $m_{3/2}$. Note that in the MSSM with BRpV, the life time would still depend on $|\lambda|^2$. In order to have a conservative idea of the corresponding bounds, we include the limits obtained in \cite{Choi:2010jt} for the point $M_1=1\TEV$ and $M_2=2M_1$. We choose this point because the three-body BR behaves similarly to our scenario. Below $10\GEV$ we consider the same reference and procedure as before.

By simple inspection of the bottom (top) panel of Fig.~(\ref{fig:tauG_BR3B_massG_M1_300}), we notice that the region $m_{3/2} > 8\,(2) \GEV$ is strongly disfavored. However, we see that masses below $10\,(2) \GEV$ are still viable. A more systematic study of these issues, considering a complete scan of the space of (effective) free parameters and complementary bounds derived from several experiments, is left for another work.

Instead, we would like to emphasize the main result of this section. We have shown that, after considering the constraints derived from the Higgs mass value and three generation neutrino observables, the space of free parameters which are relevant to investigate gravitino dark matter indirect searches in Split-SUSY with BRpV is effectively given by a 4-dimensional parameter space (note that the original parameter space is 14-dimensional). Indeed, without loss of generality, we can define the effective parameter space as ($m_{3/2}$, $M_1$, $M_2$, $\tan\beta$). 

During the last stages of our research we have realized that the gravitino life time could indeed be bounded from above for each gravitino mass, which would be an interesting result in order to impose more general constraints to our model. For instance, in the regime $M_1\simeq M_2$ and $\tan\beta > 10$ we have found that the absolute maximum always exists, and it is obtained for $M_1 \simeq m_{3/2}$. A general study of this issue is under preparation. 

\begin{figure}[!ht*]
\begin{center}
\includegraphics[width=0.8\textwidth,angle=0]{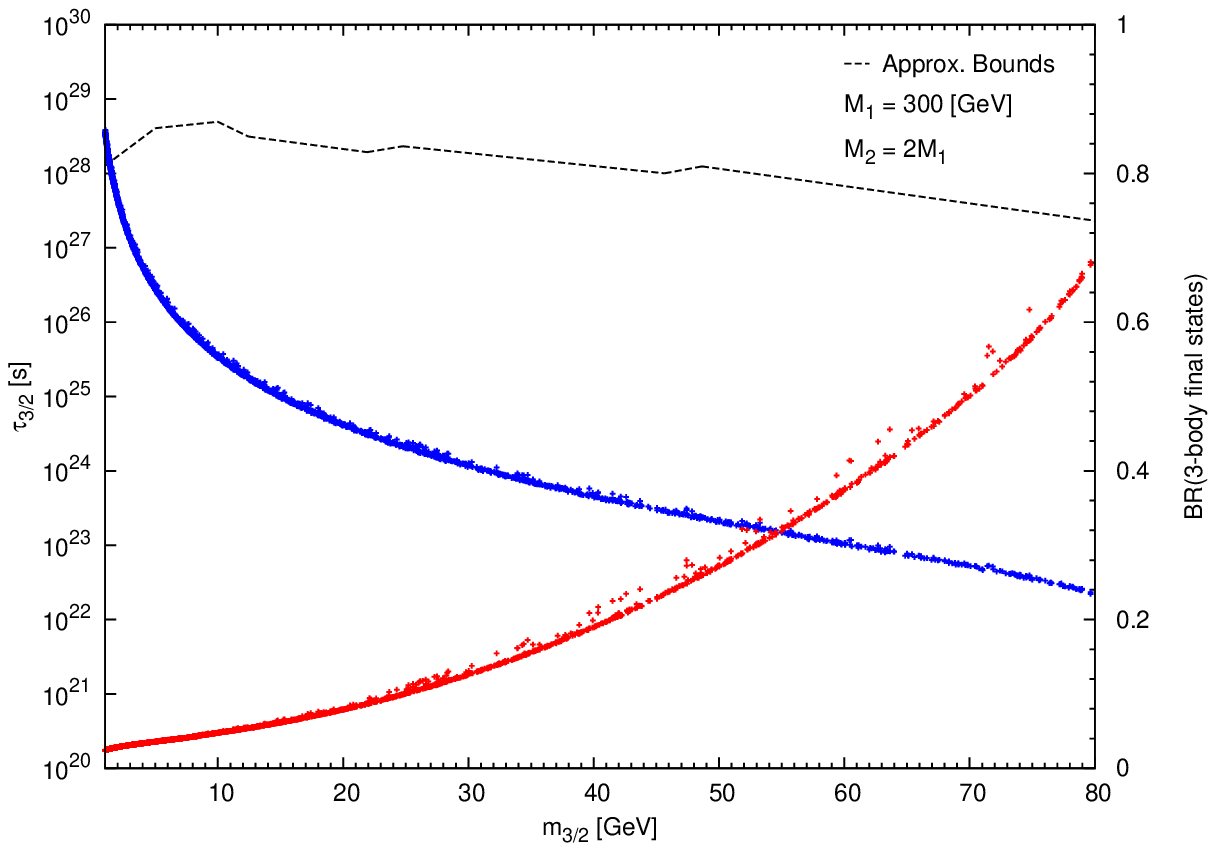} \\
\includegraphics[width=0.8\textwidth,angle=0]{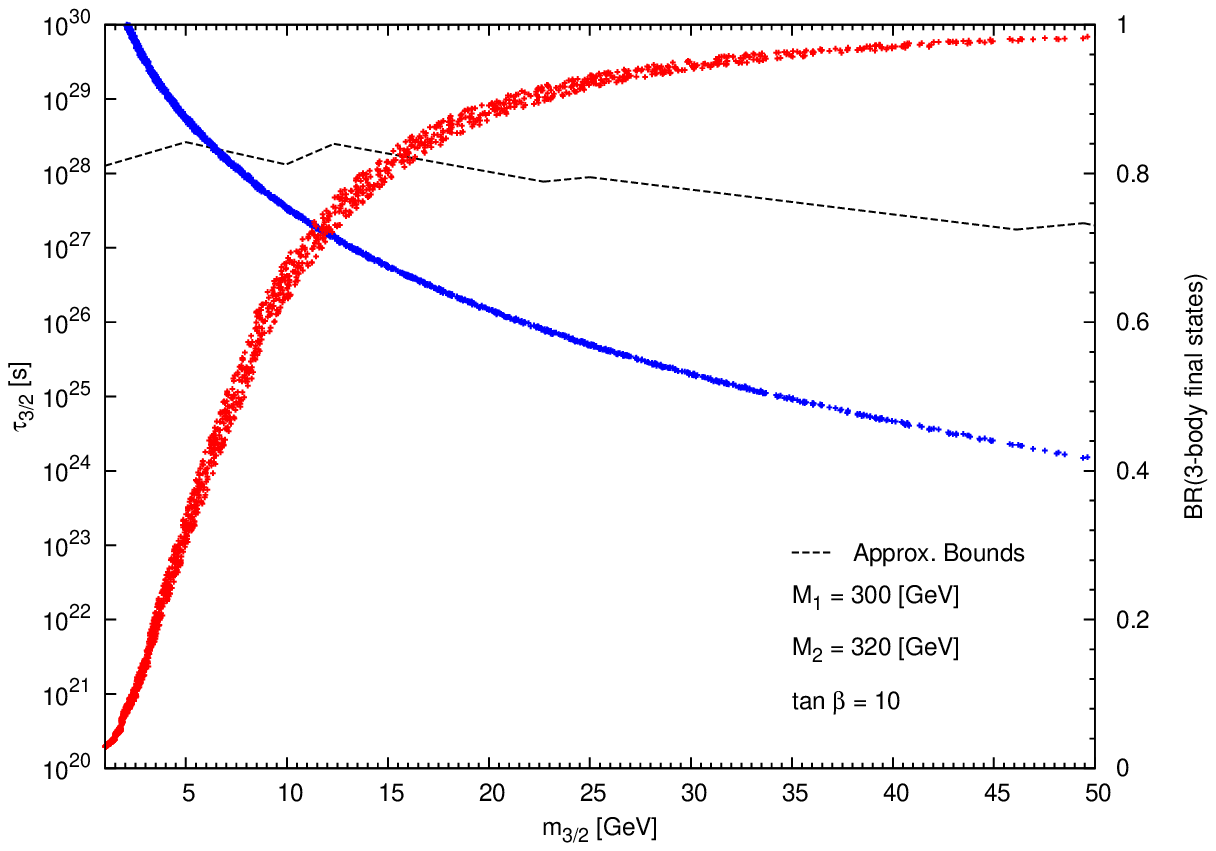}
\caption{Gravitino life time (blue) and 3-body BR (red) as a function of $m_{3/2}$ in Split-SUSY with BRpV. Top: Sugra inspired benchmark. Bottom: Compressed spectrum benchmark. Above $10\GEV$ we include the bounds of \cite{Choi:2010jt} and below $10\GEV$ we consider the bounds of \cite{Albert:2014hwa}. The reduced interval of $m_{3/2}$ in the bottom panel is explained by the correlation between $m_{3/2}$, $T_R$ and $M_1$, see Fig.~(\ref{fig:TrMassG}).}
\label{fig:tauG_BR3B_massG_M1_300}
\end{center}
\end{figure}

\section{Summary}

In order to accommodate the dark matter paradigm, we consider the simplest supergravity extension of 
supersymmetry in the Split-SUSY scenario with BRpV, with the gravitino as a dark matter candidate. We 
find this model to be consistent with the measured relic density, with the Higgs boson mass, and with neutrino observables. Furthermore, the NLSP (neutralino) decays fast enough to avoid constraints from BBN. Two-body  and three-body gravitino decays are calculated, and the total decay life time is found to be larger than the age of the universe. In Split-SUSY with BRpV we find that the atmospheric neutrino mass squared difference is directly 
related to the gravitino life time. This makes the model more falsifiable because once the gravitino BR and mass
are determined, so it is its life time (this is not the case in the MSSM with BRpV).

We numerically impose that the Higgs mass is around 126 GeV. In Split-SUSY this implies that $\tan\beta$ is related to
the Split-SUSY mass scale $\widetilde m$, and found it to be relatively low, $4\,\TEV \lsim \widetilde m \lsim 10^3\,\TEV$. 
The typical values for the reheating temperature satisfy $\widetilde{m} \ll T_{R}$, which allow us to use the
standard expressions for the gravitino relic density, using Split-SUSY RGEs. Besides, BBN constraints to NLSP decays, in our case the lightest neutralino, impose limits on its life time. They are easily satisfied in Split-SUSY with BRpV, since the neutralino life time satisfies $\tau_\chi<10^{-11}$ sec. The main decay mode is via the BRpV terms $\lambda_i$, with the decay modes into gravitinos subdominant.

We include updated neutrino physics constraints, including solar and atmospheric mass squared differences, and 
mixing angles. In Split-SUSY with BRpV in three generations we explain the atmospheric mass but we need an extra 
mass term, motivated by a higher dimensional gravity operator, for the solar mass. The flavor blindness of the 
extra operator implies that the atmospheric mass is solely explained by BRpV terms. This is the origin of the 
direct relation between the neutrino atmospheric mass and the gravitino life time.

We show that the gravitino life time is sufficiently large in comparison to the age of the universe. However it 
can decay into a photon and a neutrino (two-body decay) or other charged fermions (three-body decay). Dark matter indirect searches 
are therefore essential. A complete analysis of this model needs to incorporate experimental information and Monte Carlo simulations into the constraints we have presented in this work. We notice that in our scenario, the interplay between the Higgs mass, neutrino observables and gravitino sectors reduces the parameter space that determines the gravitino life time and branching ratios. Indeed, it is a four-dimensional space which is effectively given by $M_1,M_2,\tan\beta$ and $m_{3/2}$.

In summary, a general study of Split-SUSY with BRpV and the gravitino as dark matter candidate, 
including gravitino relic density, life time, Higgs mass, neutrino physics, and NLSP decays, allows us to see internal correlations that make it a more falsifiable model.

\section*{Acknowledgments}

{\small 
We would like to thank Michael Grefe for his very helpful comments. This work was partly funded by Conicyt grant 1100837, Anillo ACT1102 and the Conicyt Magister and Doctorate Fellowship Becas Chile Programmes. GC also was funded by the postgraduate Conicyt-Chile Cambridge Scholarship. BP also was supported by the State of S\~{a}o Paulo Research Foundation (FAPESP) and the German Research Foundation (DFG). MJG also was supported by a postgraduate fellowship from CONICET. 
}

\section*{Appendix}

\begin{appendix}

\section{Monte Carlo scan}
\label{app:montecarlo}

In order to study the interplay between the Higgs, neutrino and dark matter sectors of our model we run a Monte Carlo scan considering the parameter space of Split-SUSY with BRpV plus the gravitino sector. We select those points that satisfy simultaneously some of the most relevant experimental observations regarding these three sectors. The parameters and the corresponding intervals are given in the following table

\begin{table}[ht!]
\centering
\begin{tabular}{|c|c|c|}
\hline
Parameter & Description  & Range \\
\hline
$M_{\text{ino}}$ & Common EW-scale for gaugino soft masses & $[200 \GEV,1100 \GEV] $ \\
\hline
$M_1$ & Bino mass & $M_{\text{ino}}$ \\
$M_2$ & Wino mass & $M_{\text{ino}}+[10 \GEV,100 \GEV]$ \\
$M_3$ & Gluino mass & $M_2+[10 \GEV,200 \GEV]$ \\
$\mu$ & mu parameter & $M_{\text{ino}}+[10 \GEV,100 \GEV]$ \\
\hline
$\widetilde{m}$ & Split-SUSY scale & $[10^4 \GEV, 10^{10} \GEV] $ \\
$\tan\beta$ & Ratio of Higgs expectation values at $\widetilde{m}$ & $[1,50]$ \\
\hline
$T_{R}$ & Reheating temperature & $\widetilde{m} \times [10^2 \GEV, 10^{6} \GEV]$ \\
\hline
$\mu_g$     & Effective gravitational mass parameter & $[2\times 10^{-3} \text{eV},4\times 10^{-3} \text{eV}$] \\
$\lambda_1$ & Effective BRpV parameter & $(\bar{A}/A)\times[10^{-4}\GEV,10^{-3}\GEV]$ \\
$\lambda_2$ & - &  $(\bar{A}/A) \times [10^{-3}\GEV,10^{-2}\GEV]$ \\
$\lambda_3$ & - &  $(\bar{A}/A) \times [10^{-3}\GEV,10^{-2}\GEV]$ \\
\hline
\end{tabular}
\caption{Free parameters and the corresponding intervals of the Monte Carlo scan.}
\label{tab:mcparams}
\end{table}

We choose a common scale $M_{\text{ino}}$ for gaugino masses, such that $M_1$, $M_2$ and $M_3$ get values close to $M_{\text{ino}}$. Thus, we can decouple simultaneously, at the scale $M_{\text{ino}}$, the three gauginos from RGE computations. Indeed, for RGE computations we consider the SM content below $M_{\text{ino}}$, Split-SUSY between $M_{\text{ino}}$ and $\widetilde{m}$ and MSSM above $\widetilde{m}$. Also, we consider $M_1 < M_2 < M_3$ in order to produce a neutralino NLSP. This hierarchy is also useful to avoid stringent collider constraints on the direct production of gluinos. The Split-SUSY scale $\widetilde{m}$ is chosen such that the Higgs mass is efficiently reproduced. The reheating temperature is chosen to be greater than $\widetilde{m}$ in order to use standard expressions for the gravitino relic density, see Section~\ref{sec:relicabundance}. The intervals for each $\lambda_{i}$ are indirectly obtained depending on the value of the ratio $\bar{A}/A$, where $A$ is a function of gaugino masses and couplings, see Section~\ref{sec:neutrinos}, and $\bar{A} = -1000\, \text{eV}/\GEV^2$ is an arbitrary normalization, such that in the case $A = \bar{A}$, the obtained intervals for $\lambda_i$ reproduce a good point for neutrino physics in a reasonable time. The BRpV parameters $\epsilon_{i}$, which are independent of $\lambda_{i}$, also are generated randomly but they do not affect any computation relevant for this work. 

Every point that is recorded to show the final results has to satisfy three experimental constraints in the following order. The Higgs mass is required to lie in the interval $[125 \GEV, 127\GEV]$. This constraint determines the relation between $\tan\beta$ and $\widetilde{m}$. Then, we require that current experimental values for neutrino physics given in \cite{Forero:2014bxa} are reproduced with a $95\%$ confidence level each. Finally, we require that the gravitino relic density satisfies the $68\%$ confidence level interval computed by Planck \cite{Ade:2013zuv}. The latter fixes the mass of the gravitino. The efficiency of this process of selection is roughly $60\%$ with respect to the initial generation of points.

\section{Neutralino decay}
\label{app:neutralinodecay}

In Split-SUSY with BRpV the lightest neutralino is not stable, so it is important to show that it decays
fast enough to play no relevant role in the early universe. The neutralino decays with different
supersymmetric particles as intermediaries. Nevertheless, in Split-SUSY the squarks and 
sleptons are too heavy to contribute to the decay rate. In this situation, the neutralino decays only 
via an intermediate $Z$ and $W$ gauge boson, and via the Higgs boson $H$.

If $\chi^0_1$ decays via a $Z$ boson, a relevant Feynman rule is
\begin{center}
\vspace{-50pt} \hfill \\
\begin{picture}(110,90)(0,23) 
\Photon(10,25)(50,25){3}{5}
\ArrowLine(50,25)(78,53)
\ArrowLine(78,-3)(50,25)
\Text(10,35)[]{$Z$}
\Text(90,55)[]{$F^0_i$}
\Text(90,-5)[]{$F^0_j$}
\end{picture}
$
=\,i\,\gamma^\mu\Big[O^{znn}_{Lij}\frac{(1-\gamma_5)}{2}+
O^{znn}_{Rij}\frac{(1+\gamma_5)}{2}\Big]
$
\vspace{30pt} \hfill \\
\end{center}
\vspace{10pt}
with
\begin{eqnarray}
O^{znn}_{Lij} &=& -(O^{znn}_{Rij})^*
\nonumber\\
O^{znn}_{Rij} &=& -\frac{g}{2c_W} \left(
{\cal N}^*_{i4} {\cal N}_{j4}-{\cal N}^*_{i3} {\cal N}_{j3}-\sum^3_{k=1} {\cal N}^*_{i,4+k} {\cal N}_{j,4+k}
\right)
\label{ZF0F0Coup}
\end{eqnarray}
 where $\cal N$ corresponds to the 7 $\times$ 7  neutralino/neutrino diagonalizing mass matrix, analogous as in \cite{Hirsch:2000ef} for the MSSM with BRpV. Similarly, the relevant coupling in the decay via a $W$ gauge boson is,
\begin{center}
\vspace{-50pt} \hfill \\
\begin{picture}(110,90)(0,23) 
\Photon(10,25)(50,25){3}{5}
\ArrowLine(50,25)(78,53)
\ArrowLine(78,-3)(50,25)
\Text(10,35)[]{$W$}
\Text(90,55)[]{$F^0_i$}
\Text(90,-5)[]{$F^+_j$}
\end{picture}
$
=\,i\,\gamma^\mu\Big[O^{wnc}_{Lij}\frac{(1-\gamma_5)}{2}+
O^{wnc}_{Rij}\frac{(1+\gamma_5)}{2}\Big]
$
\vspace{30pt} \hfill \\
\end{center}
\vspace{10pt}
with
\begin{eqnarray}
O^{wnc}_{Lij} &=& -g \left[ {\cal N}^*_{i2} {\cal U}_{j1} + \frac{1}{\sqrt{2}}\left(
{\cal N}^*_{i3} {\cal U}_{j2} + \sum^3_{k=1} {\cal N}^*_{i,4+k} {\cal U}_{j,2+k}
\right)\right]
\nonumber\\
O^{wnc}_{Rij} &=& -g \left[ {\cal N}_{i2} {\cal V}^*_{j1} - \frac{1}{\sqrt{2}}
{\cal N}_{i4} {\cal V}^*_{j2} \right]
\label{WF0F+Coup}
\end{eqnarray}
with $\cal U$ and $\cal V$ the 5 $\times$ 5 chargino/charged leptons diagonalizing matrices. Note that in both cases the coupling constant involved is the usual gauge coupling $g$, but it runs with the Split-SUSY RGEs. If the neutralino is heavier than 125~$\GEV$, it can relevantly decay also via a Higgs boson. The coupling is,
\begin{center}
\vspace{-50pt} \hfill \\
\begin{picture}(110,90)(0,23) 
\DashLine(10,25)(50,25){4}
\ArrowLine(50,25)(78,53)
\ArrowLine(78,-3)(50,25)
\Text(10,35)[]{$H$}
\Text(90,55)[]{$F^0_i$}
\Text(90,-5)[]{$F^0_j$}
\end{picture}
$
=\,i\,\Big[O^{hnn}_{Lij}\frac{(1-\gamma_5)}{2}+
O^{hnn}_{Rij}\frac{(1+\gamma_5)}{2}\Big]
$
\vspace{30pt} \hfill \\
\end{center}
\vspace{10pt}
with
\begin{eqnarray}
O^{hnn}_{Lij} &=& (O^{hnn}_{Rji})^*
\nonumber\\
O^{hnn}_{Rij} &=&\frac{1}{2}\Bigg[{\cal N}_{i3} \left(\tilde g'_d {\cal N}_{j1}-\tilde g_d {\cal N}_{j2}\right)+  {\cal N}_{i4} \left(\tilde g_u  {\cal N}_{j2}-\tilde g'_u  {\cal N}_{j1}\right) 
\label{HF0F0Coup}\\
&& \qquad-\sum^{3}_{k=1}a_{k} {\cal N}_{i,k+4}\left(\tilde g_d {\cal N}_{j2}-\tilde g'_{d} {\cal N}_{j1}\right)\Bigg]
+(i\leftrightarrow j)
\nonumber
\end{eqnarray}
with $a_{k}$ corresponding to the effective parameters in Eq.~(\ref{X0mixingm}).

In our numerical calculations we use the above Feynman rules. But in order to get an algebraic inside
to the situation, we use the block-diagonalization approximation. The inverse of the neutralino mass matrix in Eq.~(\ref{X0massmat}) in Split-SUSY with BRpV is,
\begin{equation}
(M_{\chi^0})^{-1} = \frac{1}{\det{M_{\chi^0}}} \left[ \begin{matrix}
I^{gg} & I^{gh} \cr I^{hg} & I^{hh}
\end{matrix} \right],
\end{equation}
where the $2\times2$ sub-matrices are equal to,
\begin{eqnarray}
I^{gg} &=& \left[ \begin{matrix}
-M_2\mu^2+\frac{1}{2}\tilde g_u\tilde g_d v^2\mu & 
\frac{1}{4}(\tilde g_u\tilde g'_d+\tilde g'_u\tilde g_d)v^2\mu \cr
\frac{1}{4}(\tilde g_u\tilde g'_d+\tilde g'_u\tilde g_d)v^2\mu & 
-M_1\mu^2+\frac{1}{2}\tilde g'_u\tilde g'_d v^2\mu
\end{matrix} \right],
\nonumber\\
I^{gh} &=& \left[ \begin{matrix}
-\frac{1}{2}\tilde g'_uvM_2\mu+\frac{1}{8}\tilde g_u(\tilde g'_u\tilde g_d-\tilde g_u\tilde g'_d)v^3 & 
\frac{1}{2}g'_dvM_2\mu+\frac{1}{8}\tilde g_d(\tilde g'_u\tilde g_d-\tilde g_u\tilde g'_d)v^3 \cr
\frac{1}{2}\tilde g_uvM_1\mu+\frac{1}{8}\tilde g'_u(\tilde g'_u\tilde g_d-\tilde g_u\tilde g'_d)v^3 & 
-\frac{1}{2}\tilde g_dvM_1\mu+\frac{1}{8}\tilde g'_d(\tilde g'_u\tilde g_d-\tilde g_u\tilde g'_d)v^3
\end{matrix} \right],
\\
I^{hh} &=& \left[ \begin{matrix}
-\frac{1}{4}(\tilde g_u^2M_1+\tilde g'^2_uM_2)v^2 & 
M_1M_2\mu-\frac{1}{4}(\tilde g_u\tilde g_dM_1+\tilde g'_u\tilde g'_dM_2)v^2 \cr
M_1M_2\mu-\frac{1}{4}(\tilde g_u\tilde g_dM_1+\tilde g'_u\tilde g'_dM_2)v^2 & 
-\frac{1}{4}(\tilde g_d^2M_1+\tilde g'^2_dM_2)v^2
\end{matrix} \right],
\nonumber
\end{eqnarray}
$I^{hg}=(I^{gh})^T$, and the determinant given by,
\begin{equation}
\det{M_{\chi^0}} = -M_1M_2\mu^2 + \frac{1}{2}v^2\mu(\tilde g_u\tilde g_dM_1+\tilde g'_u\tilde g'_dM_2)
+ \frac{1}{16}v^4(\tilde g'_u\tilde g_d-\tilde g_u\tilde g'_d)^2
\end{equation}
Using these results, the small parameters contained in the matrix $\xi=mM_{\chi^0}^{-1}$ are,
\begin{eqnarray}
\xi_{i1} &=& \frac{v}{\det{M_{\chi^0}}} \left[ \frac{1}{2} \tilde g'_dM_2\mu +
\frac{1}{8}\tilde g_d(\tilde g'_u\tilde g_d-\tilde g_u\tilde g'_d)v^2
\right] \lambda_i
\nonumber\\
\xi_{i2} &=& -\frac{v}{\det{M_{\chi^0}}} \left[ \frac{1}{2} \tilde g_dM_1\mu -
\frac{1}{8}\tilde g'_d(\tilde g'_u\tilde g_d-\tilde g_u\tilde g'_d)v^2
\right] \lambda_i
\nonumber\\
\xi_{i3} &=& \frac{v^2}{\det{M_{\chi^0}}} 
\left[ \frac{1}{4} (\tilde g_u\tilde g_dM_1+\tilde g'_u\tilde g'_dM_2) +
\frac{v^2}{16\mu} (\tilde g'_u\tilde g_d-\tilde g_u\tilde g'_d)^2
\right] \lambda_i - \frac{\epsilon_i}{\mu}
\label{xiSS}\\
\xi_{i4} &=& -\frac{v^2}{4\det{M_{\chi^0}}} (\tilde g_d^2M_1+\tilde g'^2_dM_2) \lambda_i
\nonumber
\end{eqnarray}
Notice that the term $(\tilde g'_u\tilde g_d-\tilde g_u\tilde g'_d)$ tends to zero as the Split-SUSY 
scale approches the weak scale, and in many applications can be neglected, as done in 
\cite{Diaz:2006ee}. We also use the short notation $\xi_{i1}=\xi_1\lambda_i$, $\xi_{i2}=\xi_2\lambda_i$,
$\xi_{i3}=\xi_3\lambda_i-\epsilon_i/\mu$, and $\xi_{i4}=\xi_4\lambda_i$. Using the parameters in 
Eq.~(\ref{xiSS}) we can find the effective neutrino mass matrix in Split-SUSY with BRpV given in 
Eq.~(\ref{treenumass}).

The rotation matrix that block-diagonalizes the neutralino/neutrino mass matrix, including the 
diagonalization in the neutralino sector but not in the neutrino one, is
\begin{equation}
\cal{N}=\left[\begin{array}{cc}
N & N\xi^T \\ -\xi & {\bf 1}_{3\times3}
\end{array}\right]
\label{RN}
\end{equation}
with $N$ the 4 $\times 4$ diagonalizing neutralino matrix. If we specialize $F^0_i\rightarrow\nu_i$ and $F^0_j\rightarrow\chi^0_j$ in the $ZF^0_iF^0_j$ coupling 
in Eq.~(\ref{ZF0F0Coup}), and using Eq.~(\ref{RN}) we find,
\begin{equation}
O^{z\nu\chi}_{Rij} = \frac{g}{2c_W} \left(
2\xi_{i4} N_{j4}+\xi_{i1} N_{j1}+\xi_{i2} N_{j2}
\right) \equiv \widetilde O^{z\nu\chi}_{Rj} \lambda_i
\label{tildeOznuchi}
\end{equation}

In the case of charged leptons, in Split-SUSY with BRpV the chargino/charged lepton mass matrix is given by,
\begin{equation}
{\cal M}_{\chi^+}=\left(\begin{array}{cc} M_{\chi^+} & E^T \\ 
E' & M_\ell \end{array}\right),
\end{equation}
\noindent where the chargino sub-matrix and its inverse are,
\begin{equation}
M_{\chi^+}=\left(\begin{array}{cc} M_2 & \frac{1}{\sqrt{2}}\tilde g_uv \\ 
\frac{1}{\sqrt{2}}\tilde g_dv & \mu \end{array}\right),
\qquad
M_{\chi^+}^{-1}= \frac{1}{\det{M_{\chi^+}}}
\left(\begin{array}{cc} \mu & -\frac{1}{\sqrt{2}}\tilde g_uv \\ 
-\frac{1}{\sqrt{2}}\tilde g_dv & M_2 \end{array}\right),
\end{equation}
with $\det{M_{\chi^+}}=M_2\mu-\frac{1}{2}\tilde g_u\tilde g_dv^2$. The small parameters in this case are given
by the matrix elements of $\xi_L=EM_{\chi^+}^{-1}$, and since the matrix elements in $E'$ are proportional to 
the charged lepton masses, the analogous $\xi_R\sim m_\ell\xi_L$ are usually neglected. Given that,
\begin{equation}
E = \left(\begin{array}{cc} \frac{1}{\sqrt{2}}\tilde g_da_1v & -\epsilon_1 \\ 
\frac{1}{\sqrt{2}}\tilde g_da_2v & -\epsilon_2 \\ \frac{1}{\sqrt{2}}\tilde g_da_3v & -\epsilon_3
\end{array}\right),
\end{equation}
the relevant small parameters in the charged sector are,
\begin{eqnarray}
\xi_L^{i1} &=& \frac{\tilde g_dv}{\sqrt{2}\det{M_{\chi^+}}} \lambda_i
\nonumber\\
\xi_L^{i2} &=& -\frac{\tilde g_u\tilde g_dv^2}{2\mu\det{M_{\chi^+}}} \lambda_i - \frac{\epsilon_i}{\mu}
\end{eqnarray}
The rotation matrices that block-diagonalize the chargino/charged lepton mass matrix, are
\begin{equation}
\cal U=\left[\begin{array}{cc}
U & U\xi^T_L \\ -\xi_L & {\bf 1}_{3\times3}
\end{array}\right] \,,\qquad
\cal V=\left[\begin{array}{cc}
V & {\bf 0}_{2\times3} \\ {\bf 0}_{3\times 2} & {\bf 1}_{3\times3}
\end{array}\right]
\end{equation}

If we specialize $F^0_i\rightarrow\chi^0_i$ and $F^+_j\rightarrow\ell^+_j$ in the 
$WF^0F^+$ coupling in Eq.~(\ref{WF0F+Coup}) we find,
\begin{eqnarray}
O^{w\chi\ell}_{Lij} &=& g \left[ N_{i2} \xi_L^{j1} + \frac{1}{\sqrt{2}} N_{i3} \left(
\xi_L^{j2} - \xi_{j3} \right) - \frac{1}{\sqrt{2}} \left( N_{i1} \xi_{j1}
+ N_{i2} \xi_{j2} + N_{i4} \xi_{j4} \right)
\right] \equiv \widetilde O^{w\chi\ell}_{Lj} \lambda_i
\nonumber\\
O^{w\chi\ell}_{Rij} &=& 0
\label{tildeOwchiell}
\end{eqnarray}

Now, regarding the Higgs contribution to the neutralino decay, if we specialize
the coupling in Eq.~(\ref{HF0F0Coup}) to $F^0_i\rightarrow\nu_i$ and $F^0_j\rightarrow\chi^0_j$ we get,
\begin{eqnarray}
O^{h\nu\chi}_{Lij} &=& \frac{1}{2} \bigg[ 
(\xi_{i3}-a_i) \left(\tilde g_dN_{j2}-\tilde g'_dN_{j1}\right)
-\xi_{i4}\left(\tilde g_u N_{j2}-\tilde g'_uN_{j1}\right)+
\nonumber\\ && \qquad
N_{j3} \left( \tilde g_d\xi_{i2}-\tilde g'_d\xi_{i1} \right)
- N_{j4} \left( \tilde g_u\xi_{i2}-\tilde g'_u\xi_{i1} \right)
\bigg] \equiv \widetilde O^{h\nu\chi}_{Lj} \lambda_i
\label{tildeOhchinu}
\end{eqnarray}

The results in eqs.~(\ref{tildeOznuchi}), (\ref{tildeOwchiell}), and 
(\ref{tildeOhchinu}) tells us that the neutralino decay depends only on $\lambda_i$ (and not on $\epsilon_i$).

\section{Gravitino decay}
\label{app:gravitinodecay}

The relevant gravitino-to-matter couplings in Split-SUSY with BRpV can be computed by following a top(MSSM)-down(Split-SUSY) approach. Thus, we start by considering the gravitino to matter couplings in the MSSM with BRpV scenario \cite{Grefe:2008zz}. As we are interested in the gravitino decays, we just consider vertices that include one gravitino, one neutral or charged lepton plus a gauge boson. Typically, these couplings are proportional to mixing matrix elements, which are different from zero because of the BRpV terms. Finally, we use the matching conditions defined in Eqs.~(\ref{gtildeBC}) and (\ref{eq:mca}) to recover the couplings in pure Split-SUSY with BRpV language. Thus, we obtain

\begin{eqnarray}
 C^{\mu\nu}[\gamma,\psi_{3/2},\nu_{i}] &\simeq& \left(\frac{-i}{4M_{P}}\right)[\slashed{p},\gamma^{\mu}]\gamma^{\nu}U_{\tilde{\gamma}\nu_i}  \nonumber \\
 C^{\mu\nu}[Z,\psi_{3/2},\nu_{i}] &\simeq& \left(\frac{-i}{4M_{P}}\right)\left(-[\slashed{p},\gamma^{\mu}]\gamma^{\nu}U_{\tilde{Z}\nu_i} + P_{R}\gamma^{\mu}\gamma^{\nu}\frac{\tilde{g}_d}{\mu} \lambda_i v \right) \\
 C^{\mu\nu}[W,\psi_{3/2},\ell_i] &\simeq& \left(\frac{-i}{4M_{P}}\right)\left(-[\slashed{p},\gamma^{\mu}]\gamma^{\nu}U_{\tilde{W}\ell_i} + \sqrt{2}P_{R}\gamma^{\mu}\gamma^{\nu}\frac{\tilde{g}_d}{\mu}\lambda_i v \right), \nonumber
\label{eq:SScouplings}
\end{eqnarray}

\noindent where $M_{P}=2.4\times 10^{18}\GEV$ is the reduced Planck mass. The mixing matrix elements can be computed analytically by using the small lepton mass approximation, i.e. when the neutralino/neutrino and chargino/charged lepton BRpV mixing terms are much smaller than soft masses and the $\mu$ term, thus

\begin{eqnarray}
 U_{\tilde{\gamma}\nu_i} & \simeq & \dfrac{\mu}{2 \text{det} M_{\chi_0}^{SS}} (\tilde{g}_d M_1 s_W - \tilde{g}'_d M_2 c_W ) \lambda_i v \nonumber \\
 U_{\tilde{Z}\nu_i} & \simeq & - \dfrac{\mu}{2 \text{det} M_{\chi^0}^{SS}} (\tilde{g}_d M_1 c_W + \tilde{g}'_d M_2 s_W) \lambda_i v  \\
 U_{\tilde{W}\ell_i} & \simeq & - \dfrac{\tilde{g}_d}{\sqrt{2} \text{det} M_{\chi^{\pm}}^{SS}}\lambda_i v, \nonumber
\label{eq:SSmixings}
\end{eqnarray} 

\noindent where $M_{\chi^0}^{SS}$ and $M_{\chi^{\pm}}^{SS}$ are computed in \cite{Diaz:2006ee}. Although the couplings and mixing matrix elements are written in the gauge basis we have to recall that computations of the amplitudes has to be done in the mass basis, which in principle requires that some $U_{\text{PMNS}}$ factors should be included in the couplings. Fortunately, it turns out that after summing over the three families of neutrinos, these mixing matrix elements cancel out because of the unitarity of the $U_{\text{PMNS}}$ matrix. Also, it is interesting to notice that there is a simple map between the couplings and the mixing matrix elements of Split-SUSY with BRpV and Partial Split-SUSY (MSSM) with BRpV, which is explicitly shown in appendix~\ref{app:couplingsmatch}. 

Now, by following the approach of \cite{Diaz:2011pc}, we compute the main contributions to the gravitino decay width in the region $0 < m_{3/2} < m_H$ considering two- and three-body final state channels. We use the approximation of massless final states, but we consider exact mass thresholds. As shorter life times are more challenging experimentally speaking, the approximation of massless final states and exact thresholds is a conservative approach. In order to consider different contributions systematically, we factorize the total decay width as

\begin{equation}
  \Gamma_{3/2} = \sum_{i\in U} g_{i}(m_{3/2})h_{i}(M_{1},M_{2},\mu,\tilde g_d,\tilde g'_d,\lambda_j),
\end{equation}

\noindent where by definition $g_{i}$ is a function with mass dimension and $h_{i}$ is a dimensionless function. For a fixed value of $m_{3/2}$, the functions $g_{i}$ are just numbers. This allows us to write the total gravitino decay width as a sum of constant coefficients times functions of Split-SUSY parameters. The index $i$ runs over all the possible final states considered for the gravitino decay, i.e. $U= \{\gamma\nu_j,W^*\nu_j,\gamma (Z^*)l_j\}$ where $j$ is a family index and the two-body notation $W^*\nu_j$ or $\gamma(Z^*)l_j$ implies the sum over all three-body decays for a fixed $j$. For instance, for $i=0$ we consider the contribution from the two-body decay which is given by just one channel. For the three-body decays we consider $5$ different channels, then $i$ runs from 1 to 5 for single square amplitudes. Finally, we consider $i=mn$ with $m,n=1..5$ and $m\neq n$ for mixing terms between different channels.   

The contribution of the two-body decay is given by the decay of the gravitino into a neutrino plus a photon, whose expression is given by

\begin{eqnarray}
  \Gamma_{2B} &=& g_{0}(m_{3/2})h_{0}(M_{1},M_{2},\mu,\tilde g_d,\tilde g'_d,|\lambda|^2) \hspace{2mm} \text{such that} \nonumber \\
  g_{0}(m_{3/2}) &=&  \frac{m_{3/2}^{3}}{32\pi M_{P}^2} \\
  h_{0}(M_1,M_2,\mu,\tilde{g}_d,\tilde{g}'_d) &=& \frac{\mu^{2}}{4(\text{det} M_{\chi^0}^{SS})^2}(\tilde{g}_{d}M_{1}s_{W} -\tilde{g}'_{d}M_{2}c_{W})^{2}|\lambda|^2v^2  \nonumber
\end{eqnarray}

The contributions to the three-body decay are given by the decays of the gravitino into a fermion/anti-fermion pair plus a neutrino. The intermediate channels include the virtual exchange of $\gamma,Z$ and $W^{\pm}$ bosons. Then, in general the functions $g_i$ can be written as 

\begin{equation}
g_i(m_{3/2}) = \dfrac{1}{32(2\pi)^3 m_{3/2}^3} \int \int ds dt f_i(m_{3/2},s,t),
\label{eq:generalg}
\end{equation}

\noindent where the functions $f_i$ are dimensionless and $s$ and $t$ are the usual Mandelstam variables. Following this notation, the corresponding matrix amplitudes are given by $\langle|M_i|\rangle^2 = f_{i}h_{i}$. The explicit expressions for the functions $f_i$ are given by 

\begin{eqnarray}
 f_1(m,s,t) &=& \dfrac{q_f^2}{64 M_P^2}\dfrac{T_{11}}{s^2} \nonumber \\
 f_2(m,s,t) &=& \dfrac{g^2}{64 c_W^2 M_P^2}\dfrac{T_{22}}{(s-m_Z)^2+m_Z^2\Gamma_Z^2} \nonumber \\
 f_3(m,s,t) &=& \dfrac{g^2v^2}{256 c_W^4 M_P^2} \dfrac{T_{33}}{(s-m_Z)^2 + m_Z^2 \Gamma_Z^2} \nonumber \\
 f_4(m,s,t) &=& \dfrac{g^2}{512 M_P^2}\dfrac{T_{44}}{(s-m_W)^2+m_W^2\Gamma_W^2} \nonumber \\
 f_5(m,s,t) &=& \dfrac{g^2v^2}{1024 M_P^2} \dfrac{T_{55}}{(s-m_W)^2+m_W^2 \Gamma_W^2} \nonumber \\
 f_{12}(m,s,t) &=& \dfrac{g q_f}{64 c_W M_P^2} \dfrac{(s-m_Z^2) T_{12}}{s((s-m_Z^2)^2 + m_Z^2 \Gamma_Z^2)} \nonumber \\
 f_{13}(m,s,t) &=& \dfrac{g q_f v}{64 c_W^2 M_P^2} \dfrac{(s-m_Z^2) T_{13}}{s((s-m_Z^2)^2 + m_Z^2 \Gamma_Z^2)} \\
 f_{23}(m,s,t) &=& \dfrac{g^2 v}{64 c_W^3 M_P^2} \dfrac{T_{23}}{(s-m_Z^2)^2 + m_Z^2 \Gamma_Z^2} \nonumber \\
 f_{45}(m,s,t) &=& \dfrac{\sqrt{2}g^2 v}{512 M_P^2} \dfrac{T_{45}}{(s-m_W^2)^2 + m_W^2 \Gamma_W^2} \nonumber \\
 f_{14}(m,s,t) &=& \dfrac{\sqrt{2}g q_f}{128 M_P^2} \dfrac{ (m^2 - (s+t) - m_W^2) T_{14} }{ s((m^2-(s+t)-m_W^2)^2  + m_W^2 \Gamma_W^2 )} \nonumber \\
 f_{24}(m,s,t) &=& \dfrac{\sqrt{2}g^2 }{128 c_W M_P^2} \dfrac{[(s-m_Z^2)(m^2 - (s+t) -m_W^2 ) + m_Z m_W \Gamma_Z \Gamma_W]T_{24}}{ [(s-m_Z^2)^2 + m_Z^2 \Gamma_Z^2 ] [ (m^2-(s+t) - m_W^2)^2 + m_W^2 \Gamma_W^2 ] } \nonumber \\
 f_{34}(m,s,t) &=& \dfrac{ \sqrt{2}g^2v }{256 c_W^2 M_P^2} \dfrac{[(s-m_Z^2)( m^2-(s+t)-m_W^2 ) + m_Z m_W \Gamma_Z \Gamma_W]T_{34}}{ [(s-m_Z^2)^2 +m_Z^2 \Gamma_Z^2 ][ (m^2-(s+t)-m_W^2)^2 + m_W^2 \Gamma_W^2 ] } \nonumber \\
 f_{15}(m,s,t) &=& \dfrac{\sqrt{2}gvq_{f} }{256 M_P^2} \dfrac{[(s-m_Z^2)( m^2-(s+t)-m_W^2 ) + m_Z m_W \Gamma_Z \Gamma_W] T_{15}}{ [(s-m_Z^2)^2 +m_Z^2 \Gamma_Z^2 ][ (m^2-(s+t)-m_W^2)^2 + m_W^2 \Gamma_W^2 ] } \nonumber \\
 f_{25}(m,s,t) &=& \dfrac{g^2 v}{128 c_W M_P^2} \dfrac{[(s-m_Z^2)(m^2-(s+t) - m_W^2)+m_Z m_W \Gamma_Z \Gamma_W]T_{25}}{[(s-m_Z^2)^2+m_Z^2 \Gamma_Z^2][(m^2-(s+t)-m_W^2)^2 + m_W^2 \Gamma_W^2]} \nonumber \\
 f_{35}(m,s,t) &=& \dfrac{g^2 v^2}{256 c_W^2 M_P^2} \dfrac{[(s-m_Z^2)(m^2-(s+t)-m_W^2) + m_Z m_W \Gamma_Z \Gamma_W]T_{35}}{ [(s-m_Z^2)^2 +m_Z^2\Gamma_Z^2 ][ (m^2-(s+t)-m_W^2)^2 + m_W^2\Gamma_W^2 ] }  \nonumber
\end{eqnarray}

\noindent where $q_{f}$ is the charge of the fermion, such that $q_{e}=\sqrt{4\pi \alpha^2}$, that couples to the photon in the corresponding sub-diagram and the terms $T_{ij}$ are defined in the same way as \cite{Diaz:2011pc}. The corresponding functions $h_i$ are given by

\begin{eqnarray}
h_1(M_1,M_2,\mu,\tilde g_d,\tilde g'_d,\lambda_i) & = & |U_{\tilde{\gamma}\nu_{i}}|^2 \nonumber \\
h_2(M_1,M_2,\mu,\tilde g_d,\tilde g'_d,\lambda_i) & = &  |U_{\tilde{Z}\nu_{i}}|^2 \nonumber \\
h_3(M_1,M_2,\mu,\tilde g_d,\tilde g'_d,\lambda_i) & = & \dfrac{\tilde{g}_{d}^{2}|\lambda_{i}|^2 }{ \mu^2} \nonumber \\
h_4(M_1,M_2,\mu,\tilde g_d,\tilde g'_d,\lambda_i) & = &  |U_{\tilde{W}\ell_i}|^2 \nonumber \\
h_5(M_1,M_2,\mu,\tilde g_d,\tilde g'_d,\lambda_i) & = & h_3(M_1,M_2,\mu,\tilde g_d,\tilde g'_d,\lambda_i) \nonumber\\
h_{12}(M_1,M_2,\mu,\tilde g_d,\tilde g'_d,\lambda_i) & = & U_{\tilde{\gamma}\nu_i} U_{\tilde{Z}\nu_i} \nonumber \\
h_{13}(M_1,M_2,\mu,\tilde g_d,\tilde g'_d,\lambda_i) & = & \tilde{g}_{d}\dfrac{U_{\tilde{\gamma}\nu_i} \lambda_i}{\mu} \\
h_{23}(M_1,M_2,\mu,\tilde g_d,\tilde g'_d,\lambda_i) & = & \tilde{g}_{d}\dfrac{U_{\tilde{Z}\nu_i} \lambda_i}{\mu} \nonumber \\
h_{45}(M_1,M_2,\mu,\tilde g_d,\tilde g'_d,\lambda_i) & = & \tilde{g}_{d}\dfrac{U_{\tilde{W}\ell'_i} \lambda_i}{\mu} \nonumber \\
h_{14}(M_1,M_2,\mu,\tilde g_d,\tilde g'_d,\lambda_i) & = & U_{\tilde{\gamma}\nu_i} U_{\tilde{W}\ell'_i}  \nonumber \\
h_{24}(M_1,M_2,\mu,\tilde g_d,\tilde g'_d,\lambda_i) & = & U_{\tilde{Z}\nu_i} U_{\tilde{W}\ell'_i} \nonumber \\
h_{34}(M_1,M_2,\mu,\tilde g_d,\tilde g'_d,\lambda_i) & = & h_{45}(M_1,M_2,\mu,\tilde g_d,\tilde g'_d,\lambda_i) \nonumber \\
h_{15}(M_1,M_2,\mu,\tilde g_d,\tilde g'_d,\lambda_i) & = & h_{13}(M_1,M_2,\mu,\tilde g_d,\tilde g'_d,\lambda_i) \nonumber \\
h_{25}(M_1,M_2,\mu,\tilde g_d,\tilde g'_d,\lambda_i) & = & h_{23}(M_1,M_2,\mu,\tilde g_d,\tilde g'_d,\lambda_i) \nonumber \\
h_{35}(M_1,M_2,\mu,\tilde g_d,\tilde g'_d,\lambda_i) & = & h_{3}(M_1,M_2,\mu,\tilde g_d,\tilde g'_d,\lambda_i) \nonumber
\end{eqnarray}

Notice that each function $h_i$ is proportional to $\lambda_i^2$. After adding the contributions of the three families of neutrinos we get that each $h_i$ becomes proportional to $|\lambda|^2$.

\section{Relations between low-energy parameters}
\label{app:couplingsmatch}

It is direct to show that the relevant mixing terms that are used in the computations of the gravitino decay both in the MSSM with BRpV, Partial Split-SUSY with BRpV and Split-SUSY with BRpV are related through the following map,

\begin{table}[ht!]
\centering
\begin{tabular}{ccccc}
  MSSM with BRpV & $\leftrightarrow$ & Partial Split-SUSY with BRpV & $\leftrightarrow$ & Split-SUSY with BRpV \\ 
  $\Lambda_{i}/v_{d}$ & $\leftrightarrow$ & $\Lambda_{i}/v_{d}$ & $\leftrightarrow$ & $\lambda_{i}$  \\
  $g^{(')}\sin\beta$ & $\leftrightarrow$ & $\tilde{g}_{u}^{(')}\sin\beta$ & $\leftrightarrow$ & $\tilde{g}_{u}^{(')}$ \\
  $g^{(')}\cos\beta$ & $\leftrightarrow$ & $\tilde{g}_{d}^{(')}\cos\beta$ & $\leftrightarrow$ & $\tilde{g}_{d}^{(')}$ \\
  $v_i/v_d$ & $\leftrightarrow$ &  $b_{i}\tan\beta$ & $\leftrightarrow$ & $a_{i}$ 
\end{tabular}
\end{table}

As an example, below we explicitly show the steps in order to transform the term $\tilde{g}_{d}^{'}b_{1}v_{u}$, which is relevant for the computation of the gravitino decay to photon and neutrino in Partial Split-SUSY with BRpV \cite{Diaz:2011pc}, to the corresponding term in Split-SUSY with BRpV,

\begin{eqnarray}
 \tilde{g}_{d}^{'}b_{1}v_{u}  &=&  \tilde{g}_{d}^{'} \frac{\cos\beta}{\cos\beta}b_{1}\frac{\tan\beta}{\tan\beta}v_{u}
                                   \nonumber \\
                              &=& \tilde{g}_{d}^{'}\cos\beta b_{1} \tan\beta \frac{v_u}{\cos\beta\tan\beta} \nonumber \\
                              &\rightarrow& \tilde{g}_{d}^{'}a_{1}v,
\end{eqnarray} 

\noindent where the arrow at the third row indicates the application of the map. In the same way we can recover all the relevant couplings in the Split-SUSY with BRpV scenario. Notice that in \cite{Diaz:2011pc} it is necessary to correct the terms $g\Lambda_{i}$ by $\tilde{g}_{d}\Lambda_{i}$.

\end{appendix}

\addcontentsline{toc}{section}{References}
\bibliographystyle{utphys}
\bibliography{Cosmo_SS}

\end{document}